\documentclass[journal]{IEEEtran}

\usepackage{amssymb}
\usepackage[figuresright]{rotating}
\usepackage[tableposition=top]{caption}
\usepackage{amsmath,subfigure,epsfig,bbding}
\usepackage{epstopdf}
\usepackage{multicol}
\usepackage{booktabs}
\usepackage{multirow}
\usepackage{threeparttable}
\usepackage{color}
\usepackage{makecell}
\usepackage{amsmath}
\usepackage[linesnumbered,ruled,vlined]{algorithm2e}
\usepackage{algpseudocode}
\usepackage{amsmath}
\usepackage{graphicx}
\usepackage{array}
\newcommand{\PreserveBackslash}[1]{\let\temp=\\#1\let\\=\temp}
\newcolumntype{C}[1]{>{\PreserveBackslash\centering}p{#1}}
\newcolumntype{R}[1]{>{\PreserveBackslash\raggedleft}p{#1}}
\newcolumntype{L}[1]{>{\PreserveBackslash\raggedright}p{#1}}

\begin{document}

\title{Controllable List-wise Ranking for Universal No-reference Image Quality Assessment}

\author{Fu-Zhao~Ou,~\IEEEmembership{Student Member,~IEEE,} Yuan-Gen~Wang,~\IEEEmembership{Senior Member,~IEEE,} Jin~Li,~\IEEEmembership{Senior Member,~IEEE,}\\Guopu~Zhu,~\IEEEmembership{Senior Member,~IEEE,} and Sam~Kwong,~\IEEEmembership{Fellow,~IEEE}

\thanks{
F.-Z. Ou, Y.-G. Wang, and J. Li are with the School of Computer Science and Cyber Engineering, Guangzhou University, Guangzhou 510006, P. R. China (E-mail: oufuzhao@e.gzhu.edu.cn, wangyg@gzhu.edu.cn, lijin@gzhu.edu.cn)

G. Zhu is with the Shenzhen Institutes of Advanced Technology, Chinese Academy of Sciences, Shenzhen 518055, P. R. China (E-mail: guopu.zhu@gmail.com).

S. Kwong is with the Department of Computer Science, City University of Hong Kong, Hong Kong (E-mail: cssamk@cityu.edu.hk).
}}

\maketitle

\begin{abstract}
No-reference image quality assessment (NR-IQA) has received increasing attention in the IQA community since reference image is not always available. Real-world images generally suffer from various types of distortion. Unfortunately, existing NR-IQA methods do not work with all types of distortion. It is a challenging task to develop universal NR-IQA that has the ability of evaluating all types of distorted images. In this paper, we propose a universal NR-IQA method based on controllable list-wise ranking (CLRIQA). First, to extend the authentically distorted image dataset, we present an imaging-heuristic approach, in which the over-underexposure is formulated as an inverse of Weber-Fechner law, and fusion strategy and probabilistic compression are adopted, to generate the degraded real-world images. These degraded images are label-free yet associated with quality ranking information. We then design a controllable list-wise ranking function by limiting rank range and introducing an adaptive margin to tune rank interval. Finally, the extended dataset and controllable list-wise ranking function are used to pre-train a CNN. Moreover, in order to obtain an accurate prediction model, we take advantage of the original dataset to further fine-tune the pre-trained network. Experiments evaluated on four benchmark datasets (i.e. LIVE, CSIQ, TID2013, and LIVE-C) show that the proposed CLRIQA improves the state-of-the-art by over 9\% in terms of overall performance. The code and model are publicly available at \emph{https:$//$github.com$/$GZHU-Image-Lab$/$CLRIQA}.
\end{abstract}


\section{Introduction}

\begin{figure}[t]\centering
\setlength{\belowcaptionskip}{-0.18cm}
\subfigure[]
{\includegraphics[width=2.9in]{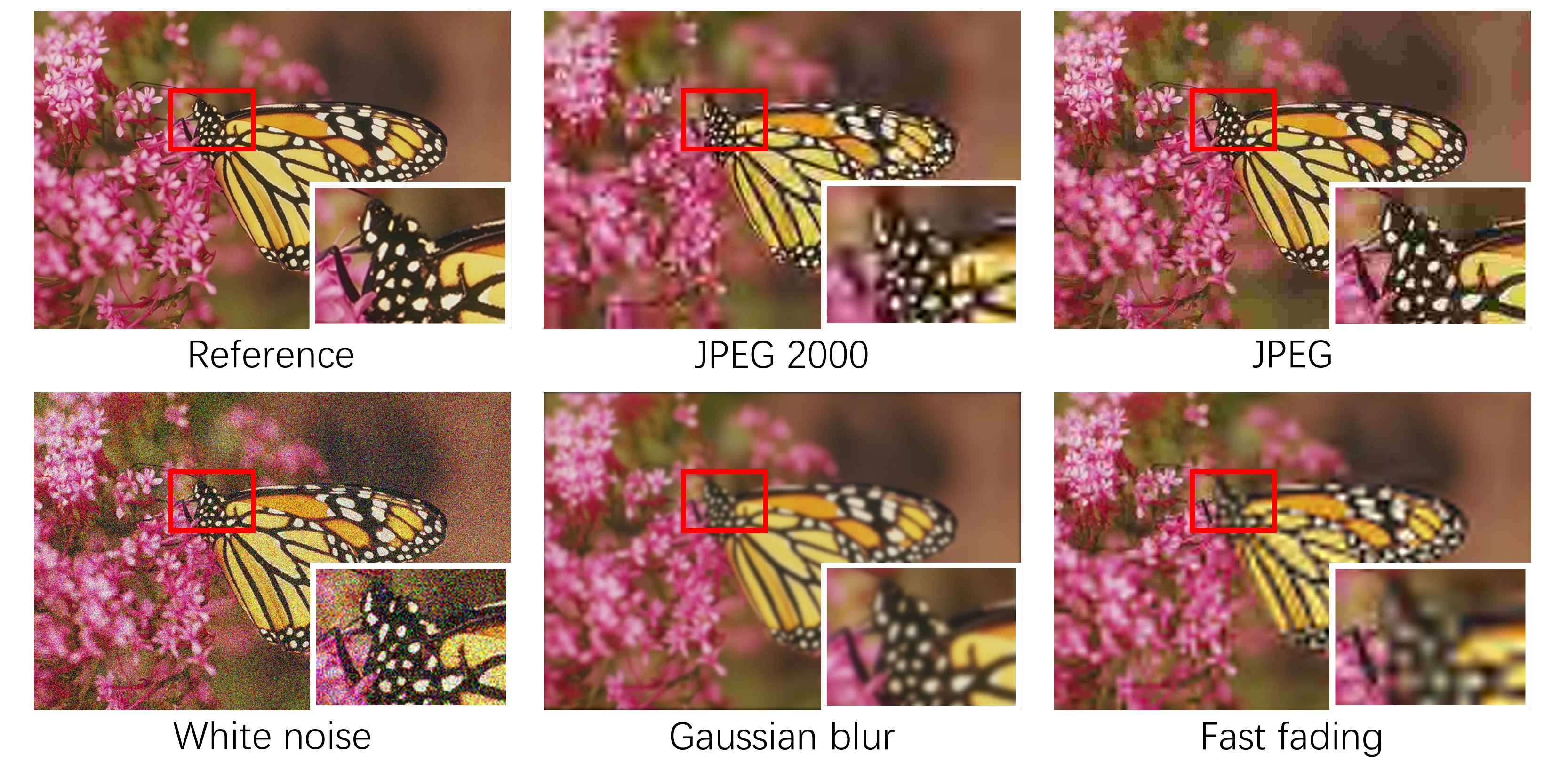}}
\subfigure[]
{\includegraphics[width=2.9in]{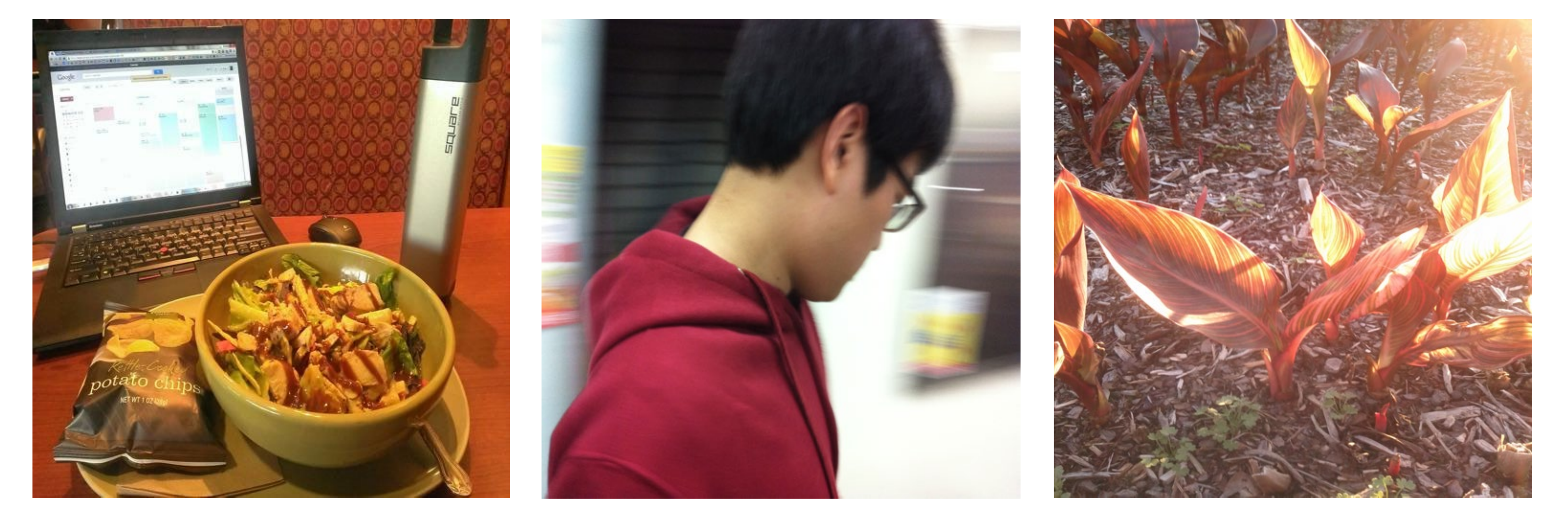}}
\caption{Illustration of synthetic and authentic distortion. (a) Synthetically distorted image. (b) Authentically distorted image.}\label{fig1}
\end{figure}

\IEEEPARstart{N}{owadays}, individuals are increasingly relying on social media. And a large part of social media interaction involves sharing images. Unfortunately, images always suffer from more or less distortion in the process of acquisition, post-processing, transmission, and storage. Accordingly, image quality assessment (IQA), which can output the quality score of a distorted image consistent with human visual system (HVS), become an important research topic \cite{Moorthy2011}. In practice, reference image is not always available. For instance, in social platform, users share images while their friends view them without any references. Hence, no-reference IQA (NR-IQA) has been the most widely and deepest studied for machine perception \cite{GMLOG2014,Yan2019}.

Commonly, images suffer from two distinct types of distortion. One is the synthetic distortion induced artificially by fast fading (FF), white noise (WN), pink Gaussian noise (PGN), JPEG (JP), JPEG2000 (JP2K), Gaussian blur (GB), global contrast decrements (CTD), and so on. Images from the LIVE \cite{LIVE}, CSIQ \cite{CSIQ2010}, and TID2013 \cite{TID2013} datasets are all the synthetically distorted ones. The other is the authentic distortion introduced inherently during capture, processing, and storage by a camera device. Each image of LIVE-C \cite{LIVE-C} is the authentically distorted one and collected without any manual post-processing. As analyzed in \cite{LIVE-C}, the authentic distortion is deemed as a mixture of overexposure, underexposure, motion-induced blur, low-light noise, and compression error, and so on. The synthetically and authentically distorted images are shown in Fig. 1. We can observe from Fig. 1(a) that it is possible to determine the type of the distorting operator, and the pristine reference might be hallucinated. Conversely, by referring to Fig. 1(b) we find that it is not well-defined distortion category for the authentically distorted image.

In literature, most NR-IQA methods are designed based on natural scene statistics (NSS) \cite{BJLC2019, NBIQA2019}. The NSS-based NR-IQA methods extract handcrafted features to build a regression model. However, the handcrafted features highly rely on the specific distortion. Hence, so far none of them can effectively deal with all types of distortion. For instance, BRISQUE \cite{BRISQUE2012} cannot handle the JP2K distortion well, GMLOG \cite{GMLOG2014} fails to evaluate the GB distortion, and while almost all of the NSS-based methods \cite{BRISQUE2012,BLIINDS22012,GMLOG2014,NFERM2015,HOSA2016,BJLC2019,NBIQA2019} perform worse with respect to the FF and CTD distortion.

Recently, the success of deep learning (DL) attracts researchers applying them to solve the NR-IQA problem \cite{Kang2014,Kim2017,PQR2018,Lin2018,Gu2019}. In general, as networks grow deeper and wider, their performances get better and better. To this end, larger and larger annotated datasets are also required for training. Unfortunately, existing IQA datasets with ground truth quality scores contain the limited number of samples, such as LIVE (808), CSIQ (896), TID2013 (3,025), and LIVE-C (1,162). However, the annotation and collection processes for IQA dataset are extremely labor-intensive and expensive. It has become a mainstream for the DL-based NR-IQA methods to enlarge the size of dataset \cite{MaK2017a}. Very recently, researchers proposed to generate the synthetically distorted images to extend the dataset \cite{USPatent2017,Liu2017,MaK2017b,Gu2019}. Although these degraded images are label-free, their quality ranking information is aware and thus can be used to train a network (i.e. learning to rank). Their designed ranking functions achieve an effective ranking, while the output scores after rank learning are largely inconsistent with the actual qualities of training images. This negatively affects the performance of trained network model. Besides, existing ranking-based NR-IQA methods do not work for LIVE-C. It is extremely challenging task to simulate the authentically distortion image in order to extend the LIVE-C dataset. This is because the authentic distortion categories are not only well-defined but complex mixture of unknown distortions \cite{LIVE-C}. What is more, it is a natural demand to design an effective ranking function to improve the performance of ranking-based NR-IQA methods.

In this paper, we develop a universal NR-IQA method based on controllable list-wise ranking. The universal function indicates that the proposed method can effectively deal with all types of distortion, including synthetic and authentic distortion. To this aim, we first present an imaging-heuristic approach to extend the authentically distorted image dataset. Moreover, inspired by the natural symmetricity assumption presented in \cite{Harsh2014}, we construct a controllable list-wise ranking function to pre-train a CNN. The extended dataset allows us to train a deeper and wider network. The designed controllable list-wise ranking function can control the output of network after sufficient training to be entirely consistent with the ground truth quality scores.

The major contributions of this paper are as follows.
\begin{itemize}
\item We present an imaging-heuristic approach to extend the LIVE-C dataset. As far as we know, LIVE-C is the unique IQA dataset that possesses authentically distorted images, and has received the most widely attention. However, up to now there has been little work that can perform well on this dataset. This is because the authentically distorted images cannot be effectively simulated by existing methods. To overcome this problem, we construct an inverse function of Weber-Fechner law to formulate over-underexposure, and then adopt fusion strategy and probabilistic compression to simulate the authentically distorted images. The size of the extended LIVE-C dataset is enlarged by over two hundred times, which greatly benefits the network train based on rank learning.

\item We design a controllable list-wise ranking function to train a CNN. Existing ranking-based NR-IQA methods, such as \cite{Gao2015,Liu2017,Liu2019,Gu2019}, cannot work well for all the IQA datasets developed so far. One of the major reasons is that these methods fail to design a powerful ranking function. Although the pair-wise ranking can achieve a good rank, the outputs of their networks are not only uncontrollable but also far inconsistent with the actual qualities of training images. To avoid this shortcoming, the controllable list-wise ranking loss limits the upper-lower bound and introduces an adaptive margin to tune the interval among ranking levels. As a result, our method significantly improves the state-of-the-art in terms of overall performance.
\end{itemize}
To our best knowledge, we are the first to propose to extend the LIVE-C dataset, and the proposed CLRIQA performs the best in implementing the universal NR-IQA function.

\section{Related Work}
In this section, we introduce the related works, including universal NR-IQA, DL-based NR-IQA, and ranking-based NR-IQA.

\subsection{Universal NR-IQA}
The universal NR-IQA method should have the ability of evaluating any images regardless of the types of distortion, datasets, and availability of reference. During the past six years, researchers attempted to design universal NR-IQA methods. As far as we know, Gao et al. \cite{Gao2013} are the first to propose the universal concept for NR-IQA. Unlike previous works that are highly dependent on the type of distortion, they combined three types of NSS models to construct a universal feature. Sang et al. \cite{Sang2014} proposed to compute a feature value without training and given distortion types. The computed value is directly acted as the quality score of an image. However, these two traditional methods are not exactly universal because they do not work with the newly discovered distortion types at all, such as the authentic distortion in the LIVE-C dataset.

\subsection{Deep learning based NR-IQA}
In recent five years, some progresses have been made in DL-based NR-IQA. Kang et al. \cite{Kang2014} are the first to apply deep CNN to NR-IQA by cropping small size of image patches for training. Tang et al. \cite{Tang2014} employed the deep belief network to extract a feature representation for NR-IQA. Kim and Lee \cite{Kim2017} used the local quality maps as intermediate targets to train a deep CNN. To achieve good consistency with human judgement, Zeng et al. \cite{PQR2018} generated five Likert-type levels to built a probabilistic quality representation (PQR), and then adopted CNN to learn the PQR of an image. Lin et al. \cite{Lin2018} and Ren et al. \cite{Ren2018} proposed to use generative adversarial network (GAN) to obtain a hallucinated reference image for NR-IQA. For integrating with different CNNs, Gu et al. \cite{Gu2018} proposed to extract features within a vector regression framework. Kim et al. \cite{DIQA2019} combined deep CNN model with two handcrafted features to further enhance the accuracy.  Except for \cite{PQR2018}, all of these DL-based NR-IQA methods cannot effectively evaluate the authentically distorted images.

\subsection{Ranking-based NR-IQA}
Ranking-based NR-IQA generally pre-trains a network by learning to rank. Gao et al. \cite{Gao2015} employs preference image pairs picked by observers to train a regression model. Ma et al. \cite{MaL2016} extract GIST features for pair-wise rank learning. Neither of these two methods is based on the deep network and dataset extension. Zhang et al. \cite{USPatent2017} is the first to propose to generate the degraded image to train a deep network according to the pair-wise rank learning. Soon after, Liu et al. \cite{Liu2017} improves \cite{USPatent2017} by introducing the Siamese network to rank learning subject to a pair-wise ranking hinge function. Gu et al. \cite{Gu2019} employs the reinforcement recursive list-wise ranking to train a CNN. A Markov decision process (MDP) is also used to implement the list-wise ranking. But regrettably, these ranking-based NR-IQA methods do not consider the extension of LIVE-C dataset for deep network training.

\section{Proposed Method}
The framework of proposed method is shown in Fig. 2. It consists of three models, which are extending datasets, pre-training, and fine-training models, respectively. In what follows, we describe each of them in detail.

\begin{figure}[t]\centering
\setlength{\belowcaptionskip}{-0.18cm}
{\includegraphics[width=3.4in]{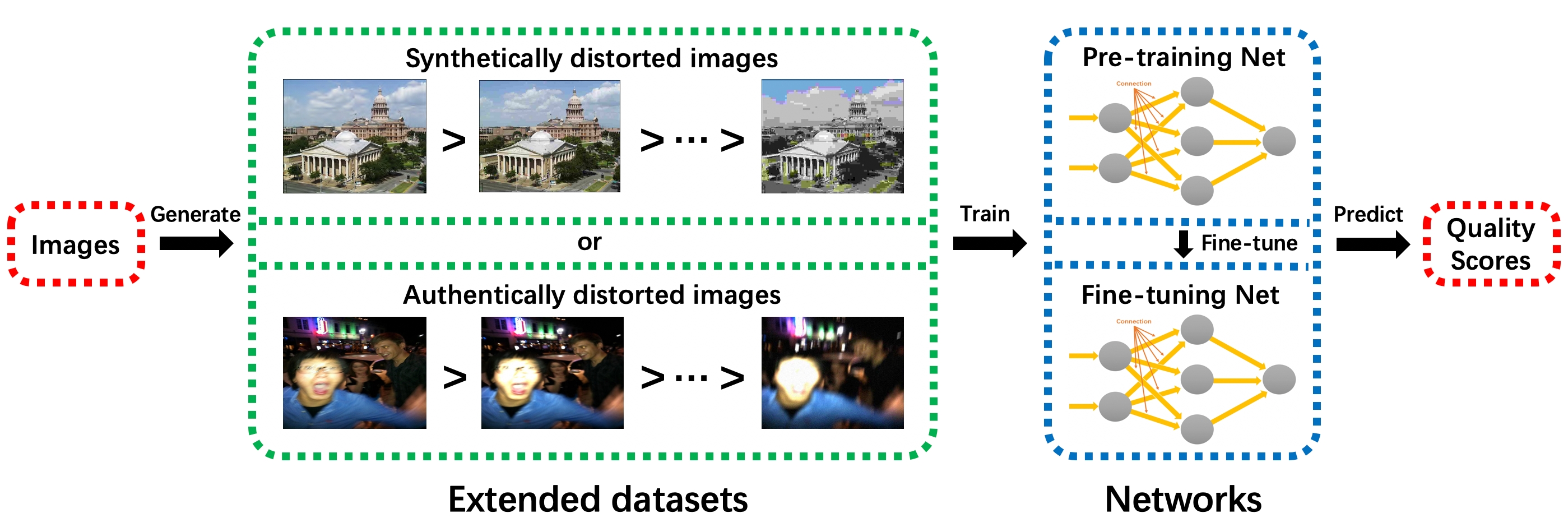}}
\caption{Overview of the proposed CLRIQA.}
\label{fig2}
\end{figure}

\subsection{\textbf{Extending Datasets}}

\subsubsection{\textbf{Extending authentically distorted image dataset}}
As analyzed in \cite{LIVE-C}, the authentic distortion can be roughly identified as a mixture of several deformations, such as overexposure, underexposure, motion-induced blur, and compression error. After a careful observation, we find that out of focus, vignetting, and contrast distortion also exist in many images of LIVE-C, too. To simulate these seven types of distortion, we present an imaging-heuristic approach to extend the authentically distorted image dataset. Specifically speaking, the value of pixels with high illumination is increased to simulate overexposure; the value of pixels with low illumination is decreased to simulate underexposure; and we deform images by using motion blur, Gaussian blur, chromatic aberrations, global contrast decrement, and JPEG compression to simulate motion-induced blur, out of focus, vignetting, contrast distortion and compression error, respectively. The detailed process is formulated as follows.

\noindent\textbf{\emph{Step 1: Mimicking over-underexposure distortion}}. Normally, the authentically distorted image is captured under highly variable illumination conditions. Weber-Fechner law shows that the change of perceptive luminance is nonlinear in HVS \cite{Van2007}. Based on the law \cite{Hecht1924}, we construct two perceptive nonlinear functions to simulate the overexposure and underexposure, respectively. For an RGB image ($I$), we first extract its luminance component, denoted as $L$. Then the overexposed luminance component ($L^{o}$) is adjusted by the following function:
\begin{equation}\label{}
L^{o}_{k}(i,j)=L(i,j)+\lambda_{1}k^{2}L(i,j)^{\gamma_{1}} + \delta_{1}kL(i,j)^{\nu_{1}},
\end{equation}
where $(i,j)$ denote the spatial indices, $k=1, 2, ..., K,$ denotes the $k$th level of $K$ distortion levels, $\lambda_{1}$, $\delta_{1}$, $\gamma_{1}$, and $\nu_{1}$ are shape parameters which are set to 6.00$\times$10$^{-3}$, 3.15$\times$10$^{-2}$, 3.02, and 2.42, respectively, by our experiment. Similarly, the underexposed luminance component ($L^{u}$) is adjusted by
\begin{equation}\label{}
L^{u}_{k}(i,j)=L(i,j)-\lambda_{2}k^{2}\log L(i,j)-{\gamma_{2}}k^{2} - \delta_{2}kL(i,j)^{\nu_{2}},
\end{equation}
where the shape parameters $\lambda_{2}$, $\gamma_{2}$, $\delta_{2}$, and $\nu_{2}$ are set to -1.8$\times$10$^{-3}$, -1.5$\times$10$^{-4}$, 2.1$\times$10$^{-5}$, and -3.01, respectively, by our experiment.
Next, we transform the adjusted luminance component back to the RGB image ($I_{t1}$).

\noindent\textbf{\emph{Step 2: Image fusion with selected operators}}. During experiments, we find that the motion-induced blur, out of focus, vignetting, and contrast distortions can be approximately simulated by the motion filter, Gaussian lowpass filter, chromatic aberration, and global contrast decrement, respectively. On the other hand, the authentic distortion is usually a many-dimensional continuum of deformations and fails to be precisely defined as the distortion categories. To better simulate such distortion, we first process $I_{t1}$ with an operator $W_{l,k}$, which is given by

\begin{equation}\label{}
I^{w}_{t1}(l,k)=W_{l,k}(I_{t1}),
\end{equation}
where $l=1,2,3,4$ denote the indices of the motion filter, Gaussian lowpass filter, chromatic aberration, and global contrast decrement operators, respectively. Here, we define a set $\Omega=\{\{1\}, \{2\}, \{3\}, \{4\}, \{1,2\}, ..., \{1,2,3,4\}\}$, which includes all possible combinations of the above four operators. Then we perform the fusion operation to obtain a fused image $I_{t2}$ as follows.
\begin{equation}\label{}
I_{t2}= \frac{1}{|\Omega(i)|}\sum_{l\in\Omega(i)}I^{w}_{t1}(l,k),\ \  i=1,2,...,15,
\end{equation}
where $|\cdot|$ denotes the set cardinality.

\noindent\textbf{\emph{Step 3: Mimicking compression errors in storage}}. The authentically distorted images is finally stored by various devices. Most devices use compression tool and thus produce the compression errors. In the final step, $I_{t2}$ is randomly selected in the probability of $1/2$ to be compressed by JPEG tool with $K$ distortion levels. Finally, the authentically distorted image is simulated, denoted by $I_{t3}$.

\textbf{Algorithm 1} describes the operation process of extending the authentically distorted image dataset. For each authentically distorted image, we can obtain 3 degraded images with respect to each distortion level $k$ after \textbf{Step 1}. These three degraded images corresponding to 3 types of distortion are overexposed, underexposed, over and underexposed ones, respectively. As done in \cite{Liu2017}, the distortion level ($K$) is also set to 5 in our method. Subsequently, all of $I_{t1}$ are sent to \textbf{Step 2} one by one. During \textbf{Step 2}, 15 types of distortion for each distortion level are generated. As a consequence, there are total 45 types of distortion. In \textbf{Step 3}, $I_{t2}$ is JPEG compressed with $K$ distortion levels. Note that for a given type of distortion, the quality of generated image gradually decreases as the distortion level $k$ increases. Finally, for each authentically distorted image, we can obtain 3$\times$15$\times$1$\times$(5+1)=270 images, where there are 3$\times$15$\times$1=45 types of distortion, 5 distortion levels, and 1 original level. Thus, the original dataset is increased to 270 times. Take LIVE-C as an example, the extended dataset will contain 1162$\times$270=313,740 images. Fig. 3(a) show the original authentically distorted images and their degraded versions. We can see from Fig. 3(a) that these degraded images seem almost the same natural as the original ones. This indicates our imaging-heuristic approach can effectively simulate the authentic distortion.

\begin{algorithm}
  \caption{Simulate Authentic Distortion}
  \KwIn{An RGB image $I$}
  \KwOut{A set of $3\times15\times5$ images $\Lambda$}
  Randomly generate $C_{tag}$ according to uniform distribution over $\{0, 1\}$;\\
  Extract the luminance component $L$ from $I$;\\
  $j=1$;\\
  \For {k=1;1 $\leq$ K}
  {
    Adjust the value of pixel with over-underexpose;\\
    Transform adjusted $L$ back to RGB image $I_{t1}$;\\
    \For {i=1;i$\leq$ 15}
    {
        \For {l=1;1$\leq$ 3}
        {
            Process with No. $l$ operator, obtain $I^{w}_{t1}(l,k)$;\\
        }
        Execute Image fusion, obtain image $I_{t2}$;\\
        \If{$C_{tag}$==1}
        {
            JPEG compress $I_{t2}$, obtain image $I_{t3}$;\\
        \textbf{else}\\
            $I_{t3} = I_{t2}$;\\
        }
        $\Lambda(j++)=I_{t3}$;
    }
  }
  Return a set of images $\Lambda$
\end{algorithm}

\subsubsection{\textbf{Extending synthetically distorted image dataset}}
Given a synthetically distorted image with a quality score, we can deform it with different types of distortion and differen distortion levels to obtain a group of worse images. This idea has been successfully implemented in previous IQA method \cite{Liu2017}. Hence, for synthetic distortion, we adopt the same method as done in \cite{Liu2017}. Fig. 3(b) show several groups of synthetically distorted images. We can see from Fig. 3(b) that all of the generated images of each group only suffer from a single type of distortion. Besides, these generated images show somewhat visual artifacts. Note that all the generated images have no ground truth score yet contain quality ranking information.

\begin{figure*}[t]\centering
\setlength{\belowcaptionskip}{-0.18cm}
\centering
\includegraphics[width=7.3in]{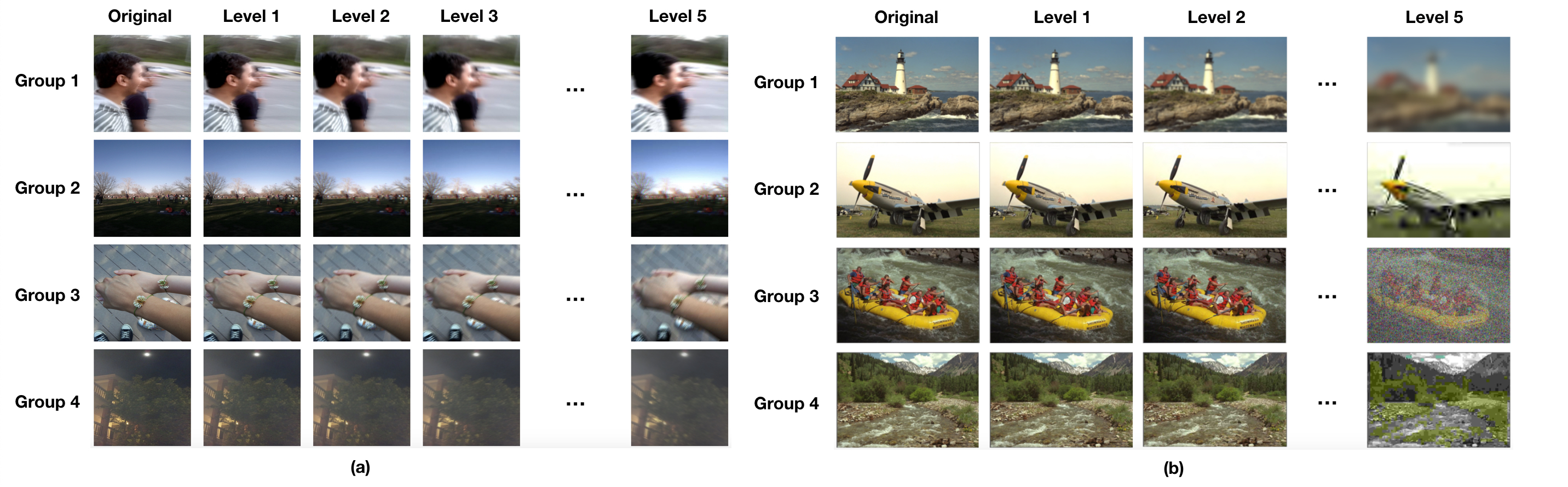}
\caption{Illustration of generated images. (a) Authentically distorted image. (b) Synthetically distorted image.}\label{fig3}
\end{figure*}

\subsection{\textbf{Pre-training Network by Rank Learning}}
After obtaining an extended dataset, we can pre-train a CNN by rank learning. We take one ground truth image and its 5 degraded versions, which are corresponding to 5 different distortion levels yet the same type of distortion, as a mini-batch. Therefore, for a mini-batch the network has 6 outputs which are passed to the loss module. Our loss function is designed as follows.
$\{I^{i}_{d} \lvert i=0, 1,..., 5\}$ denotes a mini-bitch, where $I^{0}_{d}$ denotes the ground truth image and $I^{k}_{d}$ ($k=1,...,5$) denotes the generated image with the $k$th distortion level. $y_{0}$ denotes the MOS (or DMOS) of a ground truth image (i.e., $I^{0}_{d}$). Although the ground truth quality score of the generated image is not available, we can apply the available ranking information to pre-train the network. Our ranking function is designed to
\begin{equation}\label{rank1}
\begin{split}
& \psi_{r}(I^{n}_{d}, I^{m}_{d}, y_{0}; \theta)= \\
&\sum_{n=0}^{5}\sum_{m>n}^{5}\textrm{max}(0,\phi_{\theta}(I^{n}_{d})-\phi_{\theta}(I^{m}_{d})+\frac{y_{0}}{K+1}),
\end{split}
\end{equation}
where $\theta$ denotes the network parameters and $\phi_{\theta}(\cdot)$ represents an output feature representation. Note that in Eq. (\ref{rank1}), $\frac{y_{0}}{K+1}$ is an adaptive margin with respect to the input $y_{0}$. Owing to this adaptive margin, Eq. (\ref{rank1}) can not only rank a mini-batch input in order of decreasing quality, but also control the interval among ranking levels to be $\frac{y_{0}}{K+1}$. In order to further limit the rank range, we design an upper-lower bound function:

\begin{equation}\label{rank2}
\psi_{b}(I^{0}_{d}, y_{0}; \theta) = \left\|\phi_{\theta}(I^{0}_{d})-y_{0}\right\|_{2},
\end{equation}
and
\begin{equation}\label{rank3}
\psi_{w}(I^{5}_{d}; \theta) = \begin{cases}
\left\|\phi_{\theta}(I^{5}_{d})-\tau_{w}\right\|_{2}, &\mbox{if }\phi_{\theta}(I^{5}_{d})\notin \left[ \tau_{w}, \tau_{b} \right]\\
0,  &\mbox{otherwise},
\end{cases}
\end{equation}
where $\tau_{w}$ and $\tau_{b}$ denote the worst and best quality scores in a dataset. Finally, we construct the loss function as follows.

\begin{equation}\label{loss}
\begin{split}
l_{s}(I^{0}_{d},I^{1}_{d},&...,I^{5}_{d}, y_{0};\theta)=\lambda_{r}\psi_{r}(I^{n}_{d}, I^{m}_{d}, y_{0}; \theta)\\
&+\lambda_{b}\psi_{b}(I^{0}_{d}, y_{0}; \theta)+ \lambda_{w}\psi_{w}(I^{5}_{d}; \theta),
\end{split}
\end{equation}
where $\lambda_{r}$, $\lambda_{b}$, and $\lambda_{w}$ are weight factors and $0<n<m\leq5$. Thanks to Eq. (\ref{loss}), the outputs of network can be controlled to be perfectly consistent with the actual qualities of original and degraded images. In training, the stochastic gradient descent is adopted.

\subsection{\textbf{Fine-tuning the Pre-trained Network}}
In order to obtain an efficient IQA model that can be closely in accordance with HVS, we apply the original dataset to further fine-tune the pre-trained network. In fine-tuning, each $N$ images of original dataset is taken as a batch. Then we use the Euclidean distance as the loss function, which is given by
\begin{equation}\label{}
l_{f}(I_{i},y_{i};\pi) = \frac{1}{N}\sum_{i=1}^N \left\|\phi_{\pi}(I_{i})-y_{i}\right\|_{2},
\end{equation}
where $\phi_{\pi}(I_{i})$ with network parameter $\pi$ and $y_{i}$ denote the predicted quality score and ground truth quality score of $i$th image, respectively.

\section{\textbf{Experimental Results and Analysis}}
In this section, we present the experimental setup and performance of the proposed method. The source code and network model can be downloaded at the following web address: \emph{https:$//$github.com$/$GZHU-Image-Lab$/$CLRIQA}.

\subsection{\textbf{Experimental Setup}}

\subsubsection{\textbf{Datasets description}}
In order to verify the effectiveness of the proposed method, we adopt four famous benchmark IQA datasets for testing. They are LIVE \cite{LIVE}, CSIQ \cite{CSIQ2010}, TID2013 \cite{TID2013}, and LIVE-C \cite{LIVE-C}. The details of these four datasets are illustrated in Table \uppercase\expandafter{\romannumeral1}. To be more specific, we give the distortion types for each dataset as follows.

\begin{itemize}
\item The LIVE dataset contains 29 reference images and 779 synthetically distorted images. The types of distortion are JPEG, JP2K, white noise, Gaussian blur, and fastfading.
\item The CSIQ dataset contains 30 reference images and 866 synthetically distorted images. The types of distortion are JPEG, JP2K, white noise, pink Gaussian noise, Gaussian blur, and global contrast decrements.
\item The TID2013 dataset contains 25 reference images and 3000 synthetically distorted images. The distortion types include additive Gaussian noise (\#1), additive white noise in color components (\#2), spatially correlated noise (\#3), masked noise (\#4), high frequency noise (\#5), impulse noise (\#6), quantization noise (\#7), Gaussian blur (\#8), image denoising (\#9), JPEG (\#10), JP2K (\#11), JPEG transmission errors (\#12), JP2K transmission errors (\#13), non eccentricity pattern noise (\#14), local block-wise distortions (\#15), mean shift (\#16), contrast change (\#17), change of color saturation (\#18), multiplicative Gaussian noise (\#19), comfort noise (\#20), lossy compression of noisy images (\#21), image color quantization with dither (\#22), chromatic aberrations (\#23), and sparse sampling and reconstruction (\#24).
\item The LIVE-C dataset contains 1,162 authentically distorted images, each of which is neither reference nor well-defined distortion category but labelled with MOS.
\end{itemize}

\begin{table}[]
\setlength{\abovecaptionskip}{-0.03cm}
\setlength{\belowcaptionskip}{-0.1cm}
\begin{center}
\caption{Details of four benchmark datasets.}
\resizebox{235pt}{30pt}{
\begin{tabular}{ccccccc}
\toprule
      IQA&  $\#$of Ref. &  $\#$of Dist. &Synthetic / & $\#$of Dist. & Score &  Score\\
Datasets&   Images   &    Images   &Authentic   &    Types   &  Types &  Range\\
\midrule
LIVE & 29& 779 & Synthetic & 5& DMOS& [0,100]\\
CSIQ & 30& 866& Synthetic &6 & DMOS& [0,1]\\
TID2013 & 25& 3000 & Synthetic &24& MOS& [0,9]\\
LIVE-C & -& 1162& Authentic &- & MOS& [0,100]\\
\bottomrule
\end{tabular}}
\end{center}
\end{table}\label{table1}

\subsubsection{\textbf{Implementation details}}
Our network is built on the Caffe framework \cite{caffe} and trained on a machine equipped with four GeForce RTX 2080 Ti GPUs. To match the network input, we crop the original image by 224 $\times$ 224 pixels for VGG-16 \cite{VGG16} and ResNet50 \cite{ResNet} on ImageNet. It is know that image resizing leads to information loss due to the interpolation or filter operation. Thus, we extract the sub-image by directly cropping the original image. In training, we randomly select a sub-image from the original image as network input. In testing, we randomly select 60 sub-images to calculate the mean of these 60 prediction scores as the final output. The initial learning rates of $10^{-4}$ and $10^{-5}$ are set for the pre-training and fine-tuning networks, respectively. The learning rate drops by a factor of 0.1 every 10 epochs. The weight factors $\lambda_{r}$, $\lambda_{b}$, and $\lambda_{w}$ are all set to 1. In the experiments, the dataset is randomly divided into 80$\%$ for training and the remaining 20$\%$ for testing. Note that, for the sake of clarity, the parameters of image degradation for extending datasets are given in the Appendix.

\subsubsection{\textbf{Performance metrics}}
We adopt two widely used metrics to measure the performance of IQA methods. One is the Spearman's rank ordered correlation coefficient (SROCC), which is used to measure the monotonic relationship between the ground truth quality and prediction scores. Given $N$ images, the SROCC is computed by
\begin{equation}\label{}
SROCC=1-\frac{\begin{matrix} 6\sum_{i=1}^N (\hat{y_{i}}-y_{i})^{2} \end{matrix}}{N(N^2-1)},
\end{equation}
where $\hat{y_{i}}$ and $y_{i}$ denote the ground truth quality and prediction scores of the $i$-th image, respectively. The other is the linear correlation coefficient (LCC), which measures the linear correlation between $\hat{y_{i}}$ and $y_{i}$. The LCC is formulated as
\begin{equation}\label{}
LCC=\frac{\begin{matrix} \sum_{i=1}^N (y_{i}-\bar{y})(\hat{y_{i}}-\hat{\bar{y}}) \end{matrix}}{\sqrt{\begin{matrix} \sum_{i=1}^N (y_{i}-\bar{y})^{2}\end{matrix}} \sqrt{\begin{matrix} \sum_{i=1}^N (\hat{y_{i}}-\hat{\bar{y}})^{2}\end{matrix}}}
\end{equation}
\noindent where $\bar{y}$ and $\hat{\bar{y}}$ denote the mean value of the ground truth quality and prediction scores, respectively.

\subsection{\textbf{Different CNN models analysis}}
To evaluate the performance of our proposed method, we train two widely used CNN models which are VGG-16  \cite{VGG16} and ResNet50 \cite{ResNet}. For further performance comparison on network depth, we also train a shallow CNN (S\_CNN) model. The structure of S\_CNN is shown in Fig. 4. The performances of the three network models on four datasets described in Table I are report in Table \uppercase\expandafter{\romannumeral2}. We can see from Table II that ResNet50 obtains the highest 0.980 SROCC and 0.981 PLCC on the entire LIVE dataset, and 0.849 SROCC and 0.862 PLCC on the entire LIVE-C dataset. Meanwhile it is observed that VGG-16 achieves the highest 0.952 SROCC and 0.958 PLCC on the entire CSIQ dataset, and 0.935 SROCC and 0.923 PLCC on the entire TID2013 dataset. The weighted averages on the four datasets are 0.910 SROCC and 0.916 PLCC for ResNet50, and 0.921 SROCC and 0.925 PLCC for VGG-16. It is clear that VGG-16 model performs the best in terms of overall performance. Therefore, the proposed method selects VGG-16 network to compare with other IQA methods.

\begin{figure}[t]\centering
\setlength{\belowcaptionskip}{-0.18cm}
{\includegraphics[width=3.25in]{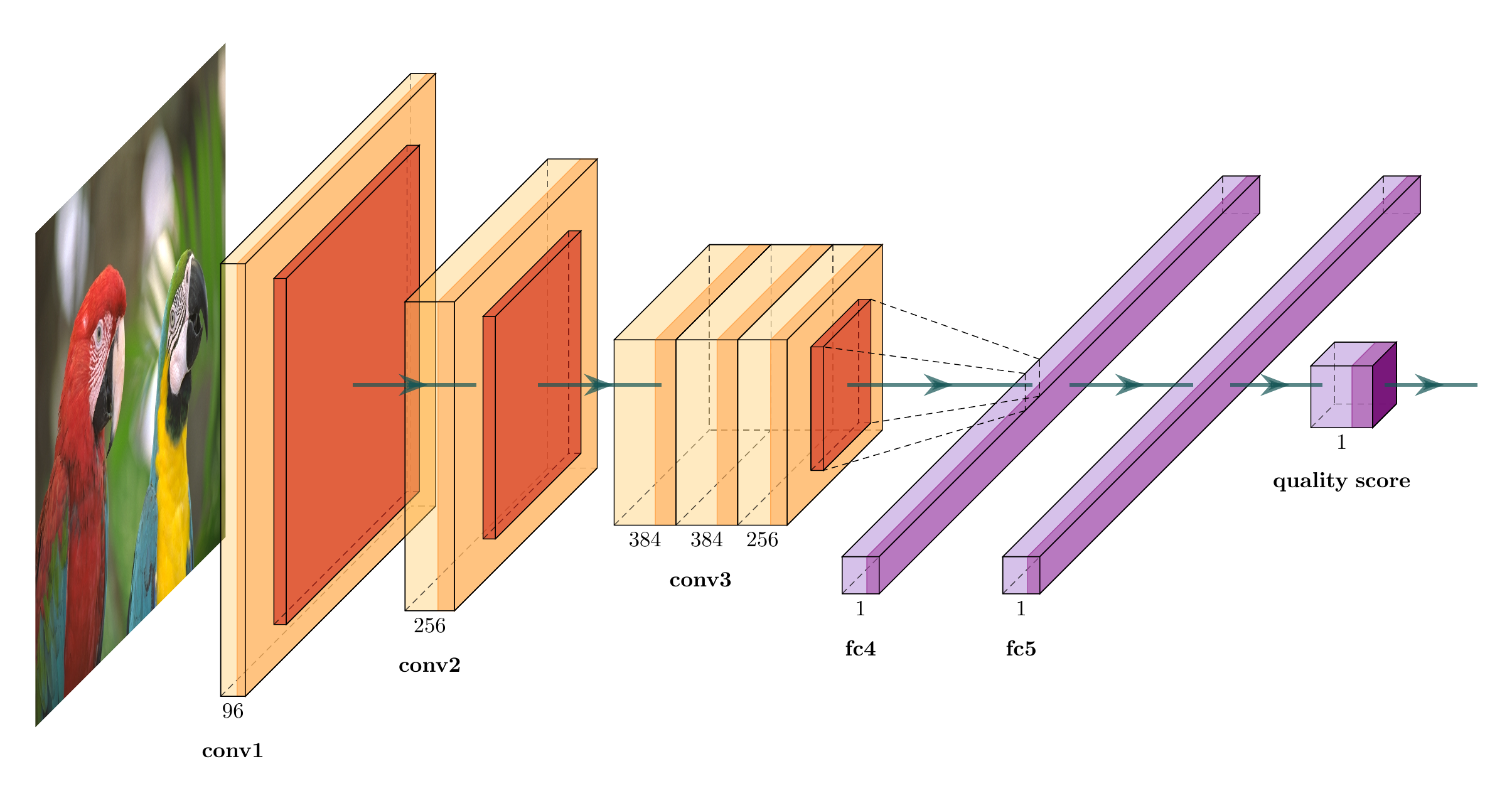}}
\caption{Structure of S\_CNN model.}
\label{fig5}
\end{figure}

\begin{table}
	\centering
	\setlength{\abovecaptionskip}{-0.03cm}
	\setlength{\belowcaptionskip}{-0.1cm}
	\caption{Performance evaluation of different CNN models through pre-training (PT) or fine-tuning after pre-training (PT+FT).}
	\begin{tabular}{|C{1cm}|C{1cm}|C{1cm}|C{1cm}|C{1cm}|C{1cm}|C{1cm}|}
		\hline
		SROCC                 & Methods   & LIVE  & CSIQ  & TID2013 & LIVE-C \\
		\hline
		
		\multirow{2}*{S\_CNN} & PT        & 0.955 & 0.789 & 0.807   & 0.761\\
		\cline{2-6}
							  & PT+FT	  & 0.963 & 0.818 & 0.840   & 0.793\\
		\hline
		\multirow{2}*{VGG-16} & PT        & 0.964 & 0.901 & 0.896 	& 0.779\\
		\cline{2-6}
							  & PT+FT	  & 0.979 & \textbf{0.953} & \textbf{0.935}   & 0.821\\
		\hline
		
		\multirow{2}*{ResNet50} & PT      & 0.966 & 0.887 & 0.838   & 0.798\\
		\cline{2-6}
							  & PT+FT	  & \textbf{0.980} & 0.919 & 0.913   & \textbf{0.849}\\
		\cline{2-6}
		
		\hline
		\hline
		\hline
		PLCC & Methods & LIVE & CSIQ & TID2013 & LIVE-C \\
		\hline
		
		\multirow{2}*{S\_CNN} & PT        & 0.930 & 0.785 & 0.783  & 0.802\\
		\cline{2-6}
							  & PT+FT	  & 0.958 & 0.812 & 0.858  & 0.836\\
		\hline
		\multirow{2}*{VGG-16} & PT        & 0.951 & 0.924 & 0.862  & 0.829\\
		\cline{2-6}
							  & PT+FT	  &0.980  & \textbf{0.958} & \textbf{0.929}  & 0.851\\
		\hline
		
		\multirow{2}*{ResNet50} & PT      & 0.968 & 0.919 & 0.845 & 0.837\\
		\cline{2-6}
							  & PT+FT	  & \textbf{0.981} & 0.931 & 0.915 & \textbf{0.862}\\
		\hline
		
	\end{tabular}
\end{table}


\subsection{\textbf{Performance Comparison with the state-of-the-art}}
The proposed CLRIQA and CLRIQA+FT is compared with 24 mainstream IQA methods on individual dataset, which are PSNR, SSIM \cite{SSIM2004}, FSIM \cite{FSIM2011}, DeepQA \cite{DeepQA2017}, NIQE \cite{NIQE2013}, IL-NIQE \cite{ILNIQE2015}, DIIVINE \cite{DIIVINE2011}, BLIINDS-\uppercase\expandafter{\romannumeral2} \cite{BLIINDS22012}, BRISQUE \cite{BRISQUE2012}, NBIQA \cite{NBIQA2019}, CORNIA \cite{CORNIA2012}, GMLOG \cite{GMLOG2014}, NFERM \cite{NFERM2015}, HOSA \cite{HOSA2016}, NRSL \cite{NRSL2016}, FRIQUEE \cite{FRIQUEE2017}, BJLC \cite{BJLC2019}, BIECON \cite{Kim2017}, H-IQA \cite{Lin2018}, PQR \cite{PQR2018}, DIQA \cite{DIQA2019}, Gao et al. \cite{Gao2015}, MT-RankIQA \cite{Liu2017, Liu2019}, and RRLRIQA\cite{Gu2019}.

\renewcommand{\arraystretch}{1}
\begin{table*}[tp]
\setlength{\abovecaptionskip}{-0.03cm}
  \centering
  \caption{Performance comparison with the state-of-the-art, including full reference IQA (FR-IQA), NSS-based NR-IQA, DL-based NR-IQA, and ranking-based NR-IQA methods. Here, the top three performers are highlighted in boldface, the symbol hyphen (-) and N/A indicate ``Absent'' and ``Not applicable'', respectively.}
  \fontsize{9.2}{9.2}\selectfont
  \begin{threeparttable}

  \setlength{\tabcolsep}{1.4mm}{
  \label{tab:performance_comparison}
    \begin{tabular}{cccccccccccc}
    \toprule
    \multirow{2}{*}{Type}&\multirow{2}{*}{Method}&
    \multicolumn{2}{c}{ LIVE (779)}&\multicolumn{2}{c}{ CSIQ (866)} &\multicolumn{2}{c}{ TID2013 (3000)} &\multicolumn{2}{c}{ LIVE-C (1162)} &\multicolumn{2}{c}{Weighted Average}\cr
    \cmidrule(lr){3-4} \cmidrule(lr){5-6} \cmidrule(lr){7-8} \cmidrule(lr){9-10} \cmidrule(lr){11-12}
     & &SROCC&PLCC&SROCC&PLCC&SROCC&PLCC&SROCC&PLCC&SROCC&PLCC\cr
    \midrule
    FR-IQA & PSNR                       &0.876&0.872&0.806&0.800&0.636&0.706&N/A&N/A&N/A&N/A\cr
      & SSIM \cite{SSIM2004}            &0.948&0.945&0.876&0.861&0.775&0.691&N/A&N/A&N/A&N/A\cr
      & FSIM \cite{FSIM2011}            &0.963&0.960&\textbf{0.931}&0.919&0.851&0.877 &N/A &N/A&N/A&N/A\cr
      & DeepQA \cite{DeepQA2017}    &\textbf{0.981}&\textbf{0.982}&\textbf{0.961}&\textbf{0.965}&\textbf{0.939}&\textbf{0.947}&N/A&N/A&N/A&N/A\cr
    \midrule
    NR-IQA (NSS)  & NIQE \cite{NIQE2013}
          &0.908&0.910&0.630&0.725&0.324&0.420&0.450&0.508&0.474&0.549\cr
      & IL-NIQE \cite{ILNIQE2015}          &0.902&0.909&0.821&0.816&0.525&0.648&0.439&0.513&0.603&0.681\cr
      & DIIVINE\cite{DIIVINE2011}
    	  &0.912&0.913&0.759&0.808&0.674&0.729&0.597&0.627&0.704&0.745\cr
      & BRISQUE \cite{BRISQUE2012}      &0.944&0.948&0.762&0.831&0.567&0.621&0.607&0.645&0.655&0.701\cr
      & BLIINDS-\uppercase\expandafter{\romannumeral2}\cite{BLIINDS22012}      &0.930&0.937&0.753&0.813&0.572&0.651&0.463&0.507&0.625&0.685\cr
      & NBIQA \cite{NBIQA2019}      &0.959&0.962&0.783&0.838&0.593&0.677&0.625&0.668&0.677&0.737\cr
      & CORNIA \cite{CORNIA2012}      &0.942&0.946&0.730&0.804&0.623&0.704&0.632&0.661&0.684&0.743\cr
      & GMLOG \cite{GMLOG2014}          &0.950&0.957&0.804&0.858&0.679&0.769&0.597&0.621&0.718&0.778\cr
      & NFERM \cite{NFERM2015}          &0.941&0.946&0.810&0.866&0.652&0.747&0.540&0.570&0.692&0.756\cr
      & HOSA \cite{HOSA2016}            &0.948&0.950&0.793&0.842&0.728&0.764&0.661&0.675&0.539&0.782\cr
      & NRSL \cite{NRSL2016}            &0.943&0.957&0.845&0.885&0.671&0.755&0.629&0.651&0.725&0.781\cr
      & FRIQUEE \cite{FRIQUEE2017}
          &0.951&0.958&0.841&0.873&0.713&0.775&0.687&0.710&0.759&0.801\cr
      & BJLC \cite{BJLC2019}            &0.956 &0.960 &0.886 &\textbf{0.918} &0.749 &0.808 &0.700 &0.732 &0.789 &0.831\cr

    \midrule
    NR-IQA (DL) & BIECON \cite{Kim2017}           &0.958&0.960&0.815&0.823&0.717&0.762&0.595&0.613&0.740&0.768\cr
      & H-IQA \cite{Lin2018}            &\textbf{0.982}&\textbf{0.982}&0.885&0.910&0.879&\textbf{0.880}&N/A  &N/A &N/A  &N/A\cr
     & PQR \cite{PQR2018}    &0.965&0.971&0.873&0.901&0.740&0.798&\textbf{0.857}&\textbf{0.882}&0.813&0.853\cr
     & DIQA \cite{DIQA2019}            &0.975&0.977&0.884&0.915&0.825&0.850&0.703&0.704&\textbf{0.830}&\textbf{0.848}\cr
    \midrule
   NR-IQA (Rank) & Gao \emph{et al.} \cite{Gao2015}&0.927&-&0.855&-    &0.767&-    &N/A  &N/A  &N/A  &N/A  \cr
      & MT-RankIQA \cite{Liu2017, Liu2019}          &0.973&0.976&-    &-    &0.806&0.827&N/A  &N/A  &N/A  &N/A \cr
      & RRLRIQA \cite{Gu2019} &0.956&0.962&0.907&0.916&0.806&0.833&N/A  &N/A  &N/A  &N/A  \cr
    \midrule
      & CLRIQA (Proposed)
          &0.964&0.951&0.901&0.924&\textbf{0.896}&0.862&\textbf{0.779}&\textbf{0.829}&\textbf{0.883}&\textbf{0.877}\cr
      & CLRIQA+FT (Proposed)                &\textbf{0.979}&\textbf{0.980}&\textbf{0.953}&\textbf{0.958}&\textbf{0.935}&\textbf{0.929}&\textbf{0.821}&\textbf{0.851}&\textbf{0.921}&\textbf{0.925}\cr
    \bottomrule
    \end{tabular}}
    \end{threeparttable}
\end{table*}

\renewcommand{\arraystretch}{1}
\begin{table*}[tp]
\setlength{\abovecaptionskip}{-0.03cm}
  \centering
  \caption{SROCC comparison on the entire TID2013 dataset. The best method is highlighted in boldface.}
  \fontsize{9}{9}\selectfont
  \begin{threeparttable}
  \setlength{\tabcolsep}{1.85mm}{
  \label{tab:performance_comparison}
    \begin{tabular}{c|ccccccccccccc}

    \Xhline{0.8pt}
    \multirow{2}{*}{Method}&\multicolumn{13}{c}{ TID2013 (3000)}\cr
    \cline{2-14}
    &\#01&\#02&\#03&\#04&\#05&\#06&\#07&\#08&\#09&\#10&\#11&\#12&\#13\cr
    \hline
    BRISQUE \cite{BRISQUE2012}           &0.706&0.523&0.776&0.295&0.836&0.802&0.682&0.861&0.500&0.790&0.779&0.254&0.723\cr
    BLIINDS-\uppercase\expandafter{\romannumeral2} \cite{BLIINDS22012}           &0.714&0.728&0.825&0.358&0.852&0.664&0.780&0.852&0.754&0.808&0.862&0.251&0.755\cr
    CORNIA \cite{CORNIA2012}
       &0.341&-0.196&0.689&0.184&0.607&-0.014&0.673&0.896&0.787&0.875&0.911&0.310&0.625\cr
    GMLOG \cite{GMLOG2014}               &0.781&0.588&0.818&0.545&0.889&0.659&0.800&0.849&0.753&0.799&0.843&0.399&0.747\cr
    NFERM \cite{NFERM2015}               &0.851&0.520&0.846&0.521&0.894&0.857&0.785&0.888&0.741&0.797&0.920&0.381&0.718\cr
    HOSA \cite{HOSA2016}                 &0.853&0.625&0.782&0.368&0.905&0.775&0.810&0.892&0.870&0.893&0.932&0.747&0.701\cr
    \hline
    Gao \emph{et al.} \cite{Gao2015}     &0.764&0.727&0.505&0.664&0.736&0.732&0.768&0.818&0.742&0.873&0.908&0.105&0.408\cr
    H-IQA \cite{Lin2018}                 &0.923&\textbf{0.880}&0.945&0.673&0.955&0.810&0.855&0.832&0.957&0.914&0.624&0.460&0.782\cr
    DIQA \cite{DIQA2019}                 &0.915&0.755&0.878&0.734&0.939&0.843&0.858&0.920&0.788&0.892&0.912&0.861&0.812\cr
    \hline
    Baseline (VGG-16)                    &0.896&0.810&0.929&0.466&0.910&0.876&0.737&0.902&0.675&0.870&0.898&0.706&0.819\cr
    RankIQA \cite{Liu2017}               &0.891&0.799&0.911&0.644&0.873&0.869&0.910&0.835&0.894&0.902&0.923&0.579&0.431\cr
    RankIQA+FT \cite{Liu2017}            &0.667&0.620&0.821&0.365&0.760&0.736&0.783&0.809&0.767&0.866&0.878&0.704&0.810\cr
    MT-RankIQA+FT \cite{Liu2019}         &0.780&0.658&0.882&0.424&0.839&0.762&0.852&0.861&0.799&0.879&0.909&0.744&0.824\cr
    \Xhline{0.8pt}
    Baseline (VGG-16)                    &0.896&0.810& 0.929&0.466&0.910&0.876&0.737&0.902&0.675&0.870&0.898&0.706&0.819\cr
    CLRIQA (Proposed)  &0.948& 0.826& 0.942& \textbf{0.859}& 0.948& 0.885& 0.941& 0.955& 0.930& 0.914& 0.896& \textbf{0.879}& 0.831\cr
    CLRIQA+FT (Proposed)  &\textbf{0.950}& 0.875&\textbf{0.953}& 0.803& \textbf{0.952}& \textbf{0.888}&\textbf{0.947}&\textbf{0.970}&\textbf{0.951}&\textbf{0.946}&\textbf{0.942}& 0.876& \textbf{0.898}\cr
    \Xhline{0.8pt}
    \multirow{2}{*}{Method}&\multicolumn{13}{c}{ TID2013 (3000)}\cr
    \cline{2-14}
    &\#14&\#15&\#16&\#17&\#18&\#19&\#20&\#21&\#22&\#23&\#24&\multicolumn{2}{|c}{ALL}\cr
    \Xhline{0.8pt}
    BRISQUE \cite{BRISQUE2012}           &0.213&0.197&0.217&0.079&0.113&0.674&0.198&0.627&0.849&0.724&0.811&\multicolumn{2}{|c}{0.567}\cr
    BLIINDS-\uppercase\expandafter{\romannumeral2} \cite{BRISQUE2012}           &0.081&0.371&0.159&-0.082&0.109&0.699&0.222&0.451&0.815&0.568&0.856&\multicolumn{2}{|c}{0.550}\cr
    CORNIA \cite{CORNIA2012}           &0.161&0.096&0.008&0.423&-0.055&0.259&0.606&0.555&0.592&0.759&0.903&\multicolumn{2}{|c}{0.651}\cr
    GMLOG \cite{GMLOG2014}               &0.190&0.318&0.119&0.224&-0.121&0.701&0.202&0.664&0.886&0.648&0.915&\multicolumn{2}{|c}{0.679}\cr
    NFERM \cite{NFERM2015}               &0.176&0.081&0.238&0.056&-0.029&0.762&0.206&0.401&0.848&0.684&0.878&\multicolumn{2}{|c}{0.652}\cr
    HOSA \cite{HOSA2016}                 &0.199&0.327&0.233&0.294&0.119&0.782&0.532&0.835&0.855&0.801&0.905&\multicolumn{2}{|c}{0.728}\cr
    \hline
    Gao \emph{et al.} \cite{Gao2015}     &0.371&0.168&0.123&0.173&0.071&0.659&0.483&0.636&0.840&0.636&0.895&\multicolumn{2}{|c}{0.481}\cr
    RankIQA \cite{Liu2017}               &0.458&0.658&0.198&0.554&0.669&0.689&0.760&0.882&0.742&0.645&0.900&\multicolumn{2}{|c}{0.806}\cr
    H-IQA \cite{Lin2018}                 &0.664&0.122&0.182&0.376&0.156&0.850&0.614&0.852&0.911&0.381&0.616&\multicolumn{2}{|c}{0.879}\cr
    DIQA \cite{DIQA2019} &0.659&0.407&0.299&0.687&-0.151&0.904&0.655&0.930&0.936&0.756&0.909&\multicolumn{2}{|c}{0.825}\cr
    \hline
    Baseline (VGG-16)    &0.398&0.449&-0.003&0.822& 0.595&0.831&0.850&0.933&0.877&0.506&0.860&\multicolumn{2}{|c}{0.718}\cr
    RankIQA \cite{Liu2017} &0.463&\textbf{0.693}&0.321&0.657&0.622&0.845&0.609&0.891&0.788&0.727&0.768&\multicolumn{2}{|c}{0.623$\downarrow$}\cr
    RankIQA+FT \cite{Liu2017} &0.512&0.622&0.268&0.613&0.662&0.619&0.644&0.800&0.779&0.629&0.859&\multicolumn{2}{|c}{0.780$\uparrow$}\cr
    MT-RankIQA+FT \cite{Liu2019}&0.458&0.658&0.198&0.554&0.669&0.689&0.760&0.882&0.742&0.645&0.900&\multicolumn{2}{|c}{0.806$\uparrow$}\cr

    \Xhline{0.8pt}
    Baseline (VGG-16)      &0.398&0.449&-0.003&0.822& 0.595&0.831&0.850&0.933&0.877&0.506&0.860&\multicolumn{2}{|c}{0.718}\cr
    CLRIQA (Proposed)      & \textbf{0.829}& 0.602& 0.679& 0.872& 0.911& 0.914& \textbf{0.964}& 0.962& 0.870& 0.901& 0.936&\multicolumn{2}{|c}{0.896$\uparrow$}\cr
    CLRIQA+FT (Proposed)      & 0.806& 0.619& \textbf{0.765}& \textbf{0.921}& \textbf{0.924}& \textbf{0.937}& 0.959& \textbf{0.969}& \textbf{0.917}& \textbf{0.905}& \textbf{0.958}&\multicolumn{2}{|c}{\textbf{0.935}$\uparrow$}\cr
    \Xhline{0.8pt}
    \end{tabular}}
    \end{threeparttable}
\end{table*}

\begin{figure*}\centering
\setlength{\belowcaptionskip}{-0.18cm}
\centering
\includegraphics[width=7.2in]{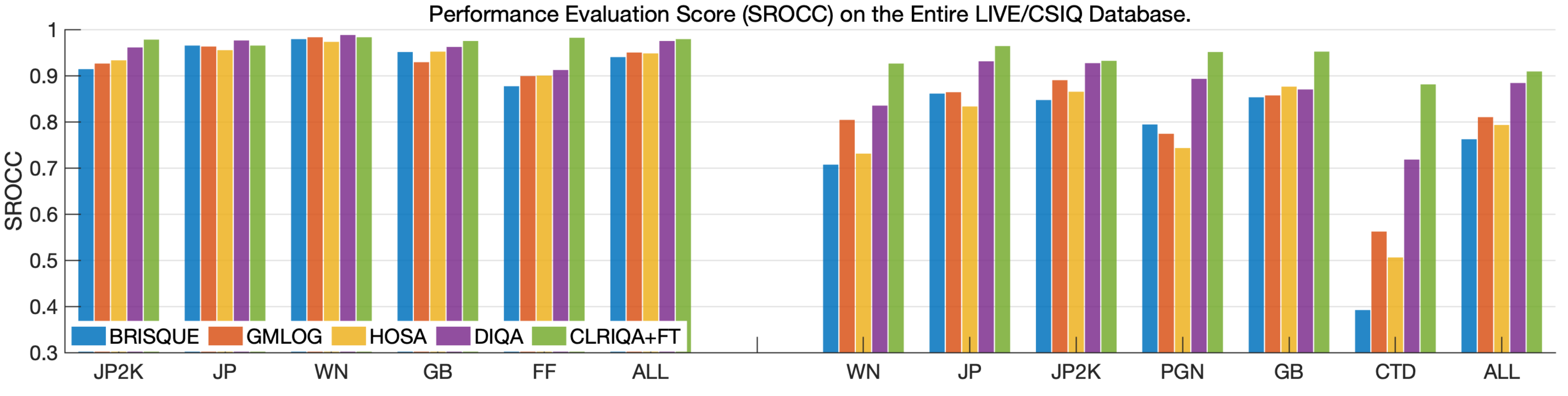}
\caption{SROCC comparison on the entire LIVE and CSID datasets. The first
five and last seven bar groups are from the LIVE and CSID datasets, respectively.}\label{fig3}
\end{figure*}

\subsubsection{\textbf{Overall performance on individual dataset}}
The comparison results are shown in Table \uppercase\expandafter{\romannumeral3}. We can see from Table \uppercase\expandafter{\romannumeral3} that for the CSIQ and TID2013 datasets, the proposed method achieves the best performance among all the NR-IQA methods. Note that the FR-IQA method if designed rationally, such as DeepQA, always performs the best for the synthetically distorted image datasets among all the IQA methods. However, the FR-IQA methods are limited in practice and not applicable in the LIVE-C dataset, either. For the LIVE dataset, our method also enters the top three of NR-IQA methods. The top three are extremely close to each other. It is interesting that for the LIVE-C dataset our method performs much better than the others except PQR. This is mainly due to the fact the LIVE-C dataset is efficiently extended by the proposed imaging-heuristic approach. It is also obvious from the last column of Table \uppercase\expandafter{\romannumeral3} that for all the datasets, our method significantly improves the state-of-the-art in terms of overall performance. It is worth mentioning that for the H-IQA method \cite{Lin2018}, like \cite{Yang2019,Gu2019}, we cannot reproduce the result of H-IQA due to lack of available source code and parameters, thus, the given result in this paper is cited from their published paper.

\subsubsection{\textbf{Evaluation on LIVE and CSIQ datasets}}
Here, we perform more detailed statistics on the performance results of 10 random rounds and compute the average SROCC of each type of distortion on the entire dataset. We select 14 competitive methods to compare with the proposed CLRIQA+FT, which are PSNR, \cite{SSIM2004,FSIM2011,DeepQA2017,DIIVINE2011,BRISQUE2012,BLIINDS22012,CORNIA2012,CNN2014,GMLOG2014,SOM2015,HOSA2016,Liu2017}, and \cite{DIQA2019}. As  shown in Fig. 5, all the compared methods perform well and steady for all the distortion types in the LIVE dataset. This is because the distortion levels are clearly discriminated and these distorted images are labelled with accurate DMOS in the LIVE dataset. Even though, our method still perform the best in overall performance. Particularly, for the FF distortion, the proposed CLRIQA achieves over 7$\%$ higher than the competitors. For the relatively complex CSID dataset, we can see from the last seven bar group of Fig. 4 that the advantages of CLRIQA+FT appear rather obvious. Interestingly, the SROCC scores of CLRIQA+FT are all above 0.85 with respect to not only each type of distortion but overall performance. This reflects the remarkable stability and generality capability of the proposed CLRIQA+FT method.
\begin{table}
	\centering
	\setlength{\abovecaptionskip}{-0.03cm}
	\setlength{\belowcaptionskip}{-0.1cm}
	\caption{SROCC and PLCC evaluation on the LIVE dataset.}
	\resizebox{245pt}{115pt}{
	\begin{tabular}{|C{0.2cm}|C{1.9cm}|C{0.55cm}|C{0.55cm}|C{0.55cm}|C{0.55cm}|C{0.55cm}|C{0.55cm}|}
		\hline
		$ $   & \textbf{SROCC}    & \textbf{JP2K}   & \textbf{JP}  & \textbf{WN}  & \textbf{GB} & \textbf{FF} & \textbf{ALL} \cr
		\hline
		\multirow{4}*{\rotatebox{90}{\textbf{FR-IQA}}}
		& PSNR                            & 0.870 & 0.885 & 0.942 & 0.763 & 0.874 & 0.876 \cr
		& SSIM \cite{SSIM2004}            & 0.939 & 0.946 & 0.964 & 0.907 & 0.941 & 0.948 \cr
		& FSIM \cite{FSIM2011}            & 0.970 & 0.981 & 0.967 & 0.972 & 0.949 & 0.963 \cr
		& DeepQA \cite{DeepQA2017}        & 0.970 & 0.978 & 0.988 & 0.971 & 0.968 & 0.981 \cr
		\hline
		\multirow{6}*{\rotatebox{90}{\textbf{NR-IQA}}}
		& DIIVINE  \cite{DIIVINE2011}                          & 0.913 & 0.910 & 0.984 & 0.921 & 0.863 & 0.912 \cr
		& BLIINDS-\uppercase\expandafter{\romannumeral2}\cite{BLIINDS22012}  & 0.929 & 0.942 & 0.969 & 0.923 & 0.889 & 0.930 \cr
		& CORNIA \cite{CORNIA2012}        & 0.943 & 0.955 & 0.976 & 0.969 & 0.906 & 0.942 \cr
		& CNN \cite{CNN2014}              & 0.952 & 0.977 & 0.978 & 0.962 & 0.908 & 0.956 \cr
		& SOM \cite{SOM2015}              & 0.947 & 0.952 & 0.984 & 0.976 & 0.937 & 0.964 \cr
		& RankIQA \cite{Liu2019}          & 0.971 & \textbf{0.978} & \textbf{0.985} & 0.979 & 0.969 & 0.973 \cr
		\hline
		& \textbf{CLRIQA+FT} 			  & \textbf{0.978} & 0.960 & 0.978 & \textbf{0.983} & \textbf{0.982} & \textbf{0.979} \cr
		
		\hline
		\hline
		\hline
		
		$ $   & \textbf{PLCC}    & \textbf{JP2K}   & \textbf{JP}  & \textbf{WN}  & \textbf{GB} & \textbf{FF} & \textbf{ALL} \cr
		\hline
		\multirow{4}*{\rotatebox{90}{\textbf{FR-IQA}}}
		& PSNR                            & 0.873 & 0.876 & 0.926 & 0.779 & 0.870 & 0.872 \cr
		& SSIM \cite{SSIM2004}            & 0.921 & 0.955 & 0.982 & 0.893 & 0.939 & 0.945 \cr
		& FSIM \cite{FSIM2011}            & 0.910 & 0.985 & 0.976 & 0.978 & 0.912 & 0.960 \cr
		& DeepQA \cite{DeepQA2017}        & - & - & - & - & - & 0.982 \cr
		\hline
		\multirow{6}*{\rotatebox{90}{\textbf{NR-IQA}}}
		& DIIVINE  \cite{DIIVINE2011}     & 0.922 & 0.921 & 0.988 & 0.923 & 0.888 & 0.917 \cr
		& BLIINDS-\uppercase\expandafter{\romannumeral2}\cite{BLIINDS22012}  & 0.935 & 0.968 & 0.980 & 0.923 & 0.888 & 0.937 \cr
		& CORNIA \cite{CORNIA2012}        & 0.951 & 0.965 & 0.987 & 0.968 & 0.917 & 0.946 \cr
		& CNN \cite{CNN2014}              & 0.953 & \textbf{0.981} & 0.984 & 0.953 & 0.933 & 0.953 \cr
		& SOM \cite{SOM2015}              & 0.952 & 0.961 & \textbf{0.991} & 0.974 & 0.954 & 0.962 \cr
		& RankIQA \cite{Liu2019}          & 0.972 & 0.978 & 0.988 & 0.982 & 0.971 & 0.976 \cr
		\hline
		& \textbf{CLRIQA+FT} 			          & \textbf{0.982} & 0.979 & 0.983 & \textbf{0.985} & \textbf{0.977} & \textbf{0.980} \cr
		\hline
		
	\end{tabular}}
\end{table}

\subsubsection{\textbf{Evaluation on TID2013 dataset}}
The detailed performance on the entire TID2013 are also reported in Table \uppercase\expandafter{\romannumeral4}. We can see from Table \uppercase\expandafter{\romannumeral4} that the proposed CLRIQA+FT performs better than other methods on 18 of all 24 types of distortion. The overall performance on SROCC achieves 0.935, which is much higher than other methods. For some challenging types of distortion (\emph{i.e.} \#12 to \#18, and \#20), CLRIQA+FT achieves satisfactory result, too. In particular, for the \#12, \#14, \#17, \#16, and \#18 distortion, all the competitors fail to challenge, but our method still performs very well.

\subsection{\textbf{Ablation Study}}
In this subsection, we perform the ablation experiment to determine the contribution of each element of the proposed model. For the synthetically distorted image, we select the TID2013 dataset for testing, while for the authentically distorted image, only LIVE-C can be used for testing. It is well known that LIVE-C is a challenging IQA dataset. It is extremely difficult for most IQA methods to improve the performance on LIVE-C. In the experiment, the baseline (BL) means the plain network that trains directly on the original dataset. Here, the use of pair-wise ranking (PWR) \cite{Liu2017, Liu2019} to pre-train on the extended dataset and then fine-tune on the original dataset is called BL+PWR. The experimental results are shown in Fig. 6.

\begin{table}
	\centering
	\setlength{\abovecaptionskip}{-0.03cm}
	\setlength{\belowcaptionskip}{-0.1cm}
	\caption{Performance comparison of PWR and CLR.}
	\begin{tabular}{|C{1cm}|C{1.2cm}|C{1cm}|C{1cm}|}
	\hline
	\textbf{Metrics}               &\textbf{Methods}  & \textbf{LIVE-C}  & \textbf{TID2013} \\
	\hline
	\multirow{2}*{\textbf{SROCC}}  &PWR      & 0.715      & 0.623   \\
	\cline{2-4}
	                               &\textbf{CLR}     & \textbf{0.779}      & \textbf{0.896}   \\
	\hline
	
	\multirow{2}*{\textbf{PLCC}}   &PWR      & 0.663      & 0.566   \\
	\cline{2-4}
	                               &\textbf{CLR}      & \textbf{0.828}      & \textbf{0.862}  \\
	\hline
	\end{tabular}
\end{table}


\begin{figure}[t]\centering
\setlength{\belowcaptionskip}{-0.18cm}
{\includegraphics[width=3.25in]{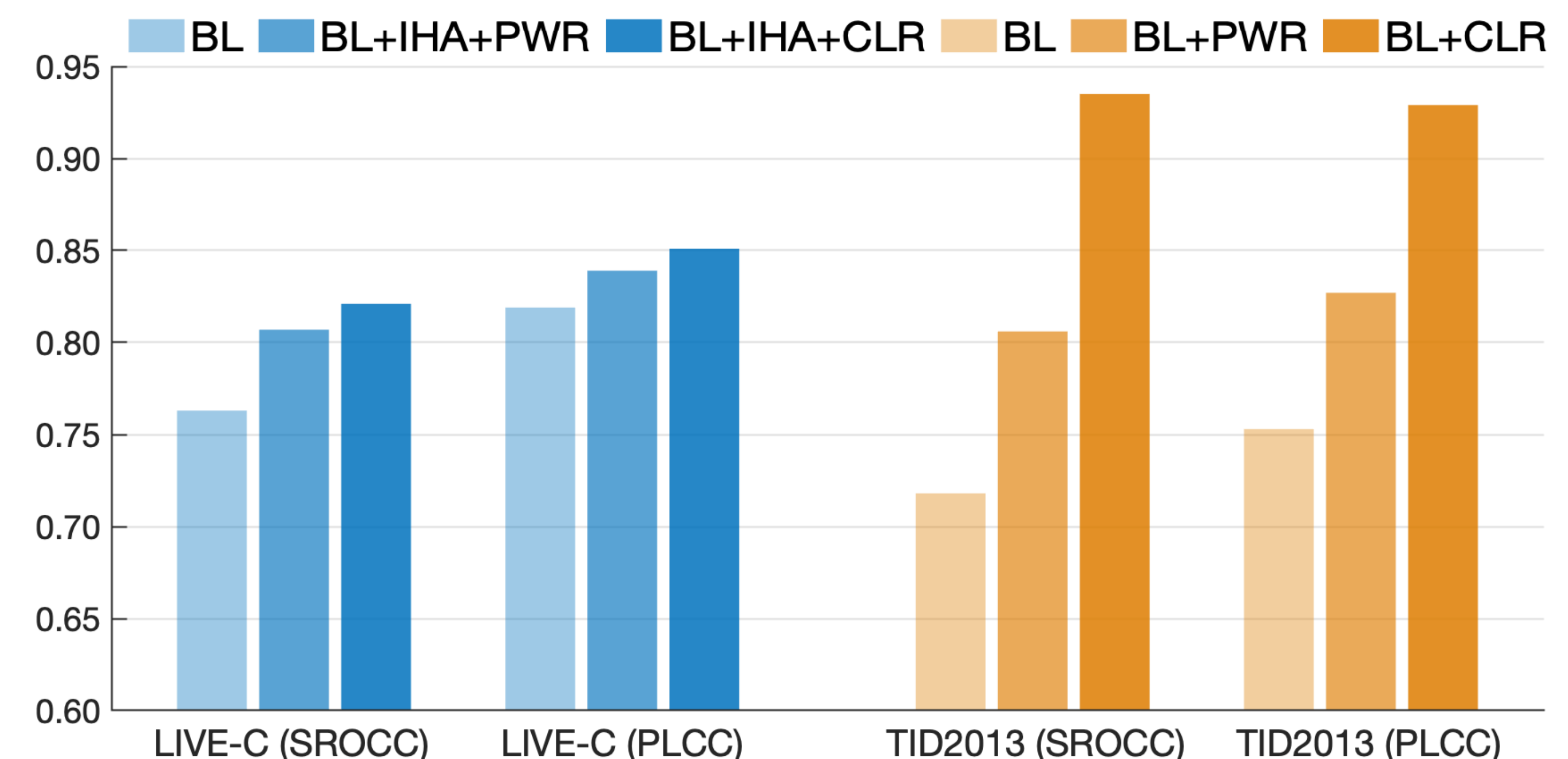}}
\caption{Ablation study on the entire LIVE-C and TID2013 datasets.}
\label{fig5}
\end{figure}

\textbf{Pre-training}. Note that the pre-training is built upon the dataset extension. In the experiment, BL+PWR obtains over 12\% and 10\% improvement on SROCC and PLCC compared with BL, respectively. This demonstrates that the dataset extension (pre-training) can greatly improve the performance. For further comparison of two pre-training methods, we also report the performance results of CLR and PWR in Table \uppercase\expandafter{\romannumeral6}.

\textbf{Imaging-heuristic approach (IHA)}. We can see from the left haft part of Fig. 6, BL+IHA+PWR is obviously higher than BL. The experimental result shows BL+IHA+PWR yields 0.807 SROCC and 0.839 PLCC, a 5.7 and 2.4\% improvement over BL, respectively. This demonstrates the dataset extension of LIVE-C (IHA) produces a significant improvement.

\textbf{Controllable list-wise ranking (CLR)}. Finally, by introducing the adaptive margin and limited upper-lower bound to the loss function, CLR achieves 0.779 SROCC and 0.828 PLCC on LIVE-C, and 0.896 SROCC and 0.862 PLCC on TID2013. Furthermore, BL+CLR obtains the highest 0.821 SROCC (1.7\% improvement over BL+PWR) and 0.851 PLCC (1.4\% improvement over BL+PWR) on LIVE-C, and 0.935 SROCC (16\% improvement over BL+PWR) and 0.929 PLCC (12\% improvement over BL+PWR) on TID2013. This further shows the efficacy of the adaptive margin and limited upper-lower bound in our loss function.

\section{Conclusions}
In this paper, we have developed a universal NR-IQA method by extending the dataset and designing a controllable list-wise ranking function. We present an imaging-heuristic approach to simulate the authentic distortion in a natural way. This is the first time, to our knowledge, that the authentically distorted images are simulated to extend the LIVE-C dataset. Furthermore, we design a controllable list-wise ranking loss function, which can effectively control the output of network to be consistent with the ground truth scores. This significantly improves the performance of trained network model. With the help of the substantially extended datasets and designed ranking loss function, our method performs the best compared with the state-of-the-art. In addition, for all the distortion types and all the datasets developed so far, the proposed method achieves the best stability and generalization ability. This indicates that the proposed CLRIQA is of a universal function method.

It should be pointed out that both the proposed method and RankIQA \cite{Liu2017, Liu2019} apply the dataset extension and rank learning, but there are significant differences between these two
methods. First, RankIQA cannot be applied to the LIVE-C dataset at all. It is known that he synthetic distortions can be easily simulated by existing image processing operations, which, however, cannot be applied to extend the authentically distortion images (such as LIVE-C). This is because the authentic distortions are considered as a complex mixture of unknown real-world deformations and thus are extremely hard to be simulated by existing methods. Therefore, RankIQA does not work with the authentically distortion images (i.e. the LIVE-C dataset). We are the first to present an approach to extend the LIVE-C dataset by constructing an inverse function of Weber-Fechner law to formulate over-underexposure and by adopting fusion strategy and probabilistic compression. By now, only the proposed method can successfully deal with the extension of LIVC-C. Second, based on a pair-wise ranking loss (PWR), RankIQA can obtain a good rank, whereas the output of its pre-training is neither controllable nor consistent with the actual quality of training image. By contrast, we design a controllable and adaptive list-wise ranking loss (CLR). The ablation study shows that the designed CLR significantly improves the PWR. As a result, the proposed method improves the state-of-the-art (include RankIQA) by over 9\% in terms of overall performance on all the four benchmark datasets.

\appendix
For each original dataset, we generate a number of degraded images with different distortion levels to extend dataset. However, the type and level of distortions depend on the original dataset. The process and related parameters are itemized in detail as follow.

\textbf{LIVE dataset}. Four widely used types of distortions including JP, JP2K, WN, and GB are used to extend LIVE dataset. For the JP, JP2K, and WN synthetic distortions, we select the different quality factors, different compression ratios, and different variances, respectively. Meanwhile, we use 2D circularly symmetric Gaussian blur kernels and change the standard deviations for the GB distorted images. The detailed parameters are reported in Tab. A1.

\renewcommand\thetable{\Alph{section}\arabic{table}}
\setcounter{table}{0}
\begin{table}[]
\setlength{\abovecaptionskip}{-0.03cm}
\setlength{\belowcaptionskip}{-0.1cm}
\begin{center}
\caption{Degradation parameters of LIVE dataset.}
\resizebox{240pt}{35pt}{
\begin{tabular}{cccccc}
\toprule
Type &  Level 1 &  Level 2 & Level 3 & Level 4 & Level 5\\
\midrule
JP & 81.6& 61.5 & 41.4 & 21.3 & 1.2\\
JP2K & 0.43& 0.33& 0.22 & 0.12 & 0.01\\
WN & 2$^{-12}$ & 2$^{-9}$ & 2$^{-6}$ &2$^{-3}$& 2$^{0}$\\
GB & 19 & 43 & 67 & 91 & 115\\
\bottomrule
\end{tabular}}
\end{center}
\end{table}\label{table5}
\setcounter{table}{0}

\textbf{CSIQ dataset}. Five types of distortions including JP, JP2K, WN, GB, and CTD are used to extend CSIQ dataset. For the first four distortion types, the generation processes are the same as used in LIVE dataset. But, the parameter for each distortion level is different. For CTD, we adjust the contrast value of colormap. The detailed generation parameters are shown in Tab. A2.

\textbf{LIVE-C dataset}. We apply the motion blur (MB) filter, GB filter, chromatic aberrations (CA), and CTD to simulate the motion-induced blur, out of focus, vignetting, and contrast distortions, respectively. Then they are randomly selected to be JPEG compressed. Those operators parameters are reported in Tab. A3.

\textbf{TID2013 dataset}. This dataset contains twenty-four types of synthetic distortions. In our experiment, we generate 17 of 24 distortion types, each of which has 5 distortion levels. The generation processes are as follows.

\begin{itemize}
\item \textbf{\#01 additive white Gaussian noise:} We adjust the variance of the Gaussian noise added in RGB color space of the target image. The variance value is set to be [0.0082, 0.019, 0.0298, 0.0406, 0.0514].
\item \textbf{\#02 additive noise in color components:} We adjust the variance of the Gaussian noise added in the YCbCr color space of the target image. The variance value is set to be [0.016, 0.025, 0.034, 0.043, 0.052].
\item \textbf{\#05 spatially correlated noise:} We first add the Gaussian noise with different variances to the Fourier domain of the target image. Then the noisy image is processed by a high-pass filter. The variance value is set to be [0.0082, 0.019, 0.0298, 0.0406, 0.0514].
\item \textbf{\#06 impulse noise:} We add the salt and pepper noise to the RGB color space of the target image. The strength value is set to be [0.008, 0.0185, 0.029, 0.0395, 0.050].
\item \textbf{\#07 quantization noise:} We set the different quantization steps for JP, the value of which is set to be [13, 10, 7, 4, 1].
\item \textbf{\#08 Gaussian blur:} We set the different quantization steps for JP, the value of which is set to be [19, 37, 55, 73, 91].
\item \textbf{\#09 image denoising:} We set the denoising value in RGB color space to be [0.008, 0.0185, 0.0290, 0.0395, 0.05].
\item \textbf{\#10 JPEG compression:} We set the quality factor that determines the DCT quantization matrix, the value of which is set to be [42, 33, 24, 15, 6].
\item \textbf{\#11 JPEG2000 compression:} We set the compression ratio, the value of which is set to be [52, 150, 343, 600].
\item \textbf{\#14 non eccentricity pattern noise:} We set the patches of size 15$\times$15, which are randomly moved to nearby regions. The number of patches is set to be [66, 120, 174, 228, 282].
\item \textbf{\#14 local blockwise distortion of different intensity:} We set image patches of 32$\times$32 are replaced by single color value. The number of patches is set to be [6, 12, 18, 24, 30].
\item \textbf{\#16 local blockwise distortion of different intensity:} We set the mean value shifting generated in both directions, the value of which is set to be [21, 30, 39, 48, 57].
\item \textbf{\#17 contrast change:} We set the contrast change generated in both directions, the value of which is set to be [0.79, 0.70, 0.61, 0.52, 0.43].
\item \textbf{\#18 change of color saturation:} We set the control factor, the value of which is set to be [0.23, -0.025, -0.28, -0.535, -0.79].
\item \textbf{\#19 multiplicative Gaussian noise:} We change the variance of the added Gaussian noise, the value of which is set to be [0.07, 0.10, 0.13, 0.16, 0.19].
\item \textbf{\#22 image color quantization with dither:} We change the quantization steps, the value of which is set to be [63, 48, 33, 18, 3].
\item \textbf{\#23 chromatic aberrations:} We adjust the mutual shifting of in the R and B channels, the value of which is set to be [4, 7, 10, 13, 16].
\end{itemize}

\setcounter{table}{1}
\begin{table}[]
\setlength{\abovecaptionskip}{-0.03cm}
\setlength{\belowcaptionskip}{-0.1cm}
\begin{center}
\caption{Degradation parameters of CSIQ dataset.}
\resizebox{240pt}{38pt}{
\begin{tabular}{cccccc}
\toprule
Type &  Level 1 &  Level 2 & Level 3 & Level 4 & Level 5\\
\midrule
JP & 42& 33 & 24 & 15 & 6\\
JP2K & 0.40& 0.31& 0.22 & 0.13 & 0.04\\
WN & 0.005 & 0.011 & 0.017 & 0.023 & 0.029\\
GB & 14 & 32 & 50 & 68 & 86\\
CTD & 0.123 & 0.207 & 0.301 & 0.395 & 0.490\\
\bottomrule
\end{tabular}}
\end{center}
\end{table}\label{table5}
\setcounter{table}{1}

\setcounter{table}{2}
\begin{table}[]
\setlength{\abovecaptionskip}{-0.03cm}
\setlength{\belowcaptionskip}{-0.1cm}
\begin{center}
\caption{The detail generation parameters of LIVE-C.}
\resizebox{240pt}{35pt}{
\begin{tabular}{cccccc}
\toprule
Type &  Level 1 &  Level 2 & Level 3 & Level 4 & Level 5\\
\midrule
MB & 6& 10.5 & 15 & 19.5 & 24\\
GB & 5& 8& 11 & 14 & 17\\
CA & 4 & 7 & 10 & 13 & 16\\
CTD & 0.11 & 0.20 & 0.29 & 0.38 & 0.47\\
JP &90	&84	&76	&67	&59\\
\bottomrule
\end{tabular}}
\end{center}
\end{table}\label{table5}
\setcounter{table}{2}

\bibliographystyle{plain} \small
\bibliography{CLRIQA}

\begin{thebibliography}{10}

\bibitem{Gao2015}
F.~{Gao}, D.~{Tao}, X.~{Gao}, and X.~{Li}.
\newblock Learning to rank for blind image quality assessment.
\newblock {\em IEEE Transactions on Neural Networks and Learning Systems},
  26(10):2275--2290, 2015.

\bibitem{Gao2013}
X.~{Gao}, F.~{Gao}, D.~{Tao}, and X.~{Li}.
\newblock Universal blind image quality assessment metrics via natural scene
  statistics and multiple kernel learning.
\newblock {\em IEEE Transactions on Neural Networks and Learning Systems},
  24(12):2013--2026, 2013.

\bibitem{LIVE-C}
D.~{Ghadiyaram} and A.~C. {Bovik}.
\newblock Massive online crowdsourced study of subjective and objective picture
  quality.
\newblock {\em IEEE Transactions on Image Processing}, 25(1):372--387, 2016.

\bibitem{FRIQUEE2017}
D.~{Ghadiyaram} and A.~C. {Bovik}.
\newblock Perceptual quality prediction on authentically distorted images using
  a bag of features approach,.
\newblock {\em Journal of Vision}, 17(1):1--25, 2017.

\bibitem{Gu2019}
J.~{Gu}, G.~{Meng}, C.~{Da}, S.~{Xiang}, and C.~{Pan}.
\newblock No-reference image quality assessment with reinforcement recursive
  list-wise ranking.
\newblock In {\em Thirty-Third AAAI Conference on Artificial Intelligence
  (AAAI)}, pages 8336--8343, 2019.

\bibitem{Gu2018}
J.~{Gu}, G.~{Meng}, J.~A. {Redi}, S.~{Xiang}, and C.~{Pan}.
\newblock Blind image quality assessment via vector regression and object
  oriented pooling.
\newblock {\em IEEE Transactions on Multimedia}, 20(5):1140--1153, 2018.

\bibitem{NFERM2015}
K.~{Gu}, G.~{Zhai}, X.~{Yang}, and W.~{Zhang}.
\newblock Using free energy principle for blind image quality assessment.
\newblock {\em IEEE Transactions on Multimedia}, 17(1):50--63, 2015.

\bibitem{Harsh2014}
P.~{Harsh} and P.~{Ravikumar}.
\newblock A representation theory for ranking functions.
\newblock In {\em Advances in Neural Information Processing Systems}, pages
  361--369, 2014.

\bibitem{ResNet}
K.~{He}, X.~{Zhang}, S.~{Ren}, and J.~{Sun}.
\newblock Deep residual learning for image recognition.
\newblock In {\em The IEEE Conference on Computer Vision and Pattern
  Recognition (CVPR)}, June 2016.

\bibitem{Hecht1924}
S.~{Hecht}.
\newblock The visual discrimination of intensity and the weber-fechner law.
\newblock {\em The Journal of general physiology}, 7(2):235--267, 1924.

\bibitem{caffe}
Y.~{Jia}, E.~{Shelhamer}, J.~{Donahue}, S.~{Karayev}, J.~{Long}, R.~{Girshick},
  S.~{Guadarrama}, and T.~{Darrell}.
\newblock Caffe: Convolutional architecture for fast feature embedding.
\newblock {\em arXiv preprint}, 1408.5093, 2014.

\bibitem{Kang2014}
L.~{Kang}, P.~{Ye}, Y.~{Li}, and D.~{Doermann}.
\newblock Convolutional neural networks for no-reference image quality
  assessment.
\newblock In {\em 2014 IEEE Conference on Computer Vision and Pattern
  Recognition (CVPR)}, pages 1733--1740, 2014.

\bibitem{CNN2014}
L.~{Kang}, P.~{Ye}, Y.~{Li}, and D.~Doermann.
\newblock Convolutional neural networks for no-reference image quality
  assessment.
\newblock In {\em The IEEE Conference on Computer Vision and Pattern
  Recognition (CVPR)}, June 2014.

\bibitem{DeepQA2017}
J.~{Kim} and S.~{Lee}.
\newblock Deep learning of human visual sensitivity in image quality assessment
  framework.
\newblock In {\em 2017 IEEE Conference on Computer Vision and Pattern
  Recognition (CVPR)}, pages 1969--1977, 2017.

\bibitem{Kim2017}
J.~{Kim} and S.~{Lee}.
\newblock Fully deep blind image quality predictor.
\newblock {\em IEEE Journal of Selected Topics in Signal Processing},
  11(1):206--220, 2017.

\bibitem{DIQA2019}
J.~{Kim}, A.~{Nguyen}, and S.~{Lee}.
\newblock Deep cnn-based blind image quality predictor.
\newblock {\em IEEE Transactions on Neural Networks and Learning Systems},
  30(1):11--24, 2019.

\bibitem{CSIQ2010}
E.~C. {Larson} and D.~M. {Chandler}.
\newblock Most apparent distortion: full-reference image quality assessment and
  the role of strategy.
\newblock {\em Journal of Electronic Imaging}, 19(1):011006:1--011006:21, 2010.

\bibitem{BJLC2019}
Q.~{Li}, W.~{Lin}, K.~{Gu}, Y.~{Zhang}, and Y.~{Fang}.
\newblock Blind image quality assessment based on joint log-contrast
  statistics.
\newblock {\em Neurocomputing}, 331:189--198, 2019.

\bibitem{NRSL2016}
Q.~{Li}, W.~{Lin}, J.~{Xu}, and Y.~{Fang}.
\newblock Blind image quality assessment using statistical structural and
  luminance features.
\newblock {\em IEEE Transactions on Multimedia}, 18(12):2457--2469, 2016.

\bibitem{Lin2018}
K.-Y. {Lin} and G.~{Wang}.
\newblock Hallucinated-iqa: No-reference image quality assessment via
  adversarial learning.
\newblock {\em IEEE Conference on Computer Vision and Pattern Recognition
  (CVPR)}, pages 732--741, 2018.

\bibitem{Liu2017}
X.~{Liu}, J.~v.~d. {Weijer}, and A.~D. {Bagdanov}.
\newblock Rankiqa: Learning from rankings for no-reference image quality
  assessment.
\newblock {\em IEEE International Conference on Computer Vision (ICCV)}, pages
  1040--1048, 2017.

\bibitem{Liu2019}
X.~{Liu}, J.~v.~d. {Weijer}, and A.~D. {Bagdanov}.
\newblock Exploiting unlabeled data in cnns by self-supervised learning to
  rank.
\newblock {\em IEEE Transactions on Pattern Analysis and Machine Intelligence},
  41(8):1862--1878, 2019.

\bibitem{MaK2017a}
K.~{Ma}, Z.~{Duanmu}, Q.~{Wu}, Z.~{Wang}, H.~{Yong}, H.~{Li}, and L.~{Zhang}.
\newblock Waterloo exploration database: New challenges for image quality
  assessment models.
\newblock {\em IEEE Transactions on Image Processing}, 26(2):1004--1016, 2017.

\bibitem{MaK2017b}
K.~{Ma}, W.~{Liu}, T.~{Liu}, Z.~{Wang}, and D.~{Tao}.
\newblock dipiq: Blind image quality assessment by learning-to-rank
  discriminable image pairs.
\newblock {\em IEEE Transactions on Image Processing}, 26(8):3951--3964, 2017.

\bibitem{MaL2016}
L.~{Ma}, L.~{Xu}, Y.~{Zhang}, Y.~{Yan}, and K.~N. {Ngan}.
\newblock No-reference retargeted image quality assessment based on pairwise
  rank learning.
\newblock {\em IEEE Transactions on Multimedia}, 18(11):2228--2237, 2016.

\bibitem{BRISQUE2012}
A.~{Mittal}, A.~K. {Moorthy}, and A.~C. {Bovik}.
\newblock No-reference image quality assessment in the spatial domain.
\newblock {\em IEEE Transactions on Image Processing}, 21(12):4695--4708, 2012.

\bibitem{NIQE2013}
A.~{Mittal}, R.~{Soundararajan}, and A.~C. {Bovik}.
\newblock Making a completely blind image quality analyzer.
\newblock {\em IEEE Signal Processing Letter}, 20(2):209--212, 2013.

\bibitem{Moorthy2011}
A.~K. {Moorthy} and A.~C. {Bovik}.
\newblock Blind image quality assessment: From natural scene statistics to
  perceptual quality.
\newblock {\em IEEE Transactions on Image Processing}, 20(12):3350--3364, 2011.

\bibitem{DIIVINE2011}
A.~K. {Moorthy} and A.~C. {Bovik}.
\newblock Blind image quality assessment: From natural scene statistics to
  perceptual quality,.
\newblock {\em IEEE Transactions on Image Processing}, 20(12):3350--3364, 2011.

\bibitem{NBIQA2019}
F.-Z. {Ou}, Y.-G. {Wang}, and G.~{Zhu}.
\newblock A novel blind image quality assessment method based on refined
  natural scene statistics.
\newblock In {\em 2019 IEEE International Conference on Image Processing
  (ICIP)}, pages 1004--1008, 2019.

\bibitem{TID2013}
N.~{Ponomarenko}, L.~{Jin}, O.~{Ieremeiev}, V.~{Lukin}, K.~{Egiazarian},
  J.~{Astola}, B.~{Vozel}, K.~{Chehdi}, M.~{Carli}, F.~{Battisti}, and C.-C.~J.
  {Kuo}.
\newblock Image database tid2013: Peculiarities, results and perspectives.
\newblock {\em Signal Processing: Image Communication}, 30:57--77, 2015.

\bibitem{Ren2018}
H.~{Ren}, D.~{Chen}, and Y.~{Wang}.
\newblock Ran4iqa: Restorative adversarial nets for no-reference image quality
  assessment.
\newblock In {\em Thirty-Second AAAI Conference on Artificial Intelligence
  (AAAI)}, pages 7308--7314, 2018.

\bibitem{BLIINDS22012}
M.~A. {Saad}, A.~C. {Bovik}, and C.~{Charrier}.
\newblock Blind image quality assessment: A natural scene statistics approach
  in the dct domain.
\newblock {\em IEEE Transactions on Image Processing}, 21(8):3339--3352, 2012.

\bibitem{Sang2014}
Q.~{Sang}, X.~{Wu}, C.~{Li}, and Y.~{Lu}.
\newblock Universal blind image quality assessment using contourlet transform
  and singular-value decomposition.
\newblock {\em Journal of Electronic Imaging}, 23(6):061104:1--061104:9, 2014.

\bibitem{LIVE}
H.~R. {Sheikh}, M.~F. {Sabir}, and A.~C. {Bovik}.
\newblock A statistical evaluation of recent full reference image quality
  assessment algorithms.
\newblock {\em IEEE Transactions on Image Processing}, 15(11):3440--3451, 2006.

\bibitem{VGG16}
K.~{Simonyan} and A.~{Zisserman}.
\newblock Very deep convolutional networks for large-scale image recognition.
\newblock {\em arXivpreprint arXiv: 1409.1556}, 2014.

\bibitem{Tang2014}
H.~{Tang}, N.~{Joshi}, and A.~{Kapoor}.
\newblock Blind image quality assessment using semi-supervised rectifier
  networks.
\newblock In {\em 2014 IEEE Conference on Computer Vision and Pattern
  Recognition (CVPR)}, pages 2877--2884, 2014.

\bibitem{Van2007}
J.~H. {van Hateren} and H.~P. {Snippe}.
\newblock Simulating human cones from mid-mesopic up to high-photopic
  luminances.
\newblock {\em Journal of Vision}, 7(4):1--11, 2007.

\bibitem{SSIM2004}
Z.~{Wang}, A.~C. {Bovik}, H.~R. {Sheikh}, and E.~P. {Simoncelli}.
\newblock Image quality assessment: from error visibility to structural
  similarity.
\newblock {\em IEEE Transactions on Image Processing}, 13(4):600--612, 2004.

\bibitem{HOSA2016}
J.~{Xu}, P.~{Ye}, Q.~{Li}, H.~{Du}, Y.~{Liu}, and D.~{Doermann}.
\newblock Blind image quality assessment based on high order statistics
  aggregation.
\newblock {\em IEEE Transactions on Image Processing}, 25(9):4444--4457, 2016.

\bibitem{GMLOG2014}
W.~{Xue}, X.~{Mou}, L.~{Zhang}, A.~C. {Bovik}, and X.~{Feng}.
\newblock Blind image quality assessment using joint statistics of gradient
  magnitude and laplacian features.
\newblock {\em IEEE Transactions on Image Processing}, 23(11):4850--4862, 2014.

\bibitem{Yan2019}
Q.~{Yan}, D.~{Gong}, and Y.~{Zhang}.
\newblock Two-stream convolutional networks for blind image quality assessment.
\newblock {\em IEEE Transactions on Image Processing}, 28(5):2200--2211, 2019.

\bibitem{Yang2019}
X.~{Yang}, F.~{Li}, and H.~{Liu}.
\newblock A comparative study of dnn-based models for blind image quality
  prediction.
\newblock In {\em 2019 IEEE International Conference on Image Processing
  (ICIP)}, pages 1019--1023, Sep. 2019.

\bibitem{CORNIA2012}
P.~{Ye}, J.~{Kumar}, L.~{Kang}, and D.~{Doermann}.
\newblock Unsupervised feature learning framework for no-reference image
  quality assessment.
\newblock pages 1098--1105, 2012.

\bibitem{PQR2018}
H.~{Zeng}, L.~{Zhang}, and A.~C. {Bovik}.
\newblock Blind image quality assessment with a probabilistic quality
  representation.
\newblock In {\em 2018 IEEE International Conference on Image Processing
  (ICIP)}, pages 609--613, 2018.

\bibitem{ILNIQE2015}
L.~{Zhang}, L.~{Zhang}, and A.~C. {Bovik}.
\newblock A feature-enriched completely blind image quality evaluator.
\newblock {\em IEEE Transactions on Image Processing}, 24(8):2579--2591, 2015.

\bibitem{FSIM2011}
L.~{Zhang}, L.~{Zhang}, X.~{Mou}, and D.~{Zhang}.
\newblock Fsim: A feature similarity index for image quality assessment.
\newblock {\em IEEE Transactions on Image Processing}, 20(8):2378--2386, 2011.

\bibitem{SOM2015}
P.~{Zhang}, W.~{Zhou}, L.~{Wu}, and H.~{Li}.
\newblock Som: Semantic obviousness metric for image quality assessment.
\newblock In {\em The IEEE Conference on Computer Vision and Pattern
  Recognition (CVPR)}, June 2015.

\bibitem{USPatent2017}
Q.~{Zhang}, Z.~{Ji}, L.~{Shi}, and I.~{Ovsiannikov}.
\newblock Label-free non-reference image quality assessment via deep neural
  network, 2017.
\newblock US Patent 9,734,567.

\end{thebibliography}

\end{document}